\documentclass[a4paper, 11pt]{article}

\usepackage{authblk}
\usepackage{inputenc}
\usepackage[T1]{fontenc}
\usepackage{cite}

\usepackage{amsthm}
\usepackage{amssymb}
\usepackage{mathtools}
\usepackage{stmaryrd}	\SetSymbolFont{stmry}{bold}{U}{stmry}{m}{n}	
\usepackage{bm}
\usepackage{mathrsfs}
\usepackage{euscript}

\usepackage{tikz}
\usepackage{tkz-euclide}
\usepackage{tikz-cd}
\usepackage{pgfplots}
\pgfplotsset{compat=1.13}

\usepackage{silence}
\usetikzlibrary{braids}

\usetikzlibrary{arrows,calc}
\usetikzlibrary{shapes}
\usetikzlibrary{positioning}
\usetikzlibrary{decorations.pathreplacing,decorations.markings,intersections}
\tikzset{
->-/.style args={#1rotate#2}{decoration={markings, mark=at position #1 with {\arrow[scale=1.5,rotate = #2 ]{stealth}}}, postaction={decorate}}
}
\tikzset{cross/.style={preaction={-,draw=white,line width=6pt}}}

\usepackage{verbatim}
\usepackage{paralist}
\usepackage{tcolorbox}

\usepackage[colorlinks]{hyperref}
\hypersetup{
	linkcolor=blue,
	citecolor=red,
	urlcolor=teal,
	pdftitle = {},
	pdfauthor = {}
}

\newcommand{\Z}{\ensuremath{\mathbb{Z}}}

\newcommand{\R}{\ensuremath{\mathbb{R}}}
\newcommand{\CC}{\ensuremath{\mathbb{C}}}

\newcommand{\Aut}{\operatorname{Aut}}
\newcommand{\Tr}{\operatorname{tr}}

\newcommand{\tr}{\ensuremath{\mathrm{tr}}}

\newcommand{\identity}{\ensuremath{\mathrm{id}}}

\newcommand{\Hom}{\operatorname{Hom}}
\newcommand{\Obj}{\operatorname{Obj}}

\newcommand{\End}{\operatorname{End}}

\newcommand{\ket}[1]{\left|#1\right\rangle}
\newcommand{\bra}[1]{\left\langle#1\right|}

\theoremstyle{plain}
\newtheorem{proposition}{Proposition}

\theoremstyle{definition}
\newtheorem{definition}[proposition]{Definition}
\newtheorem{definition-theorem}[proposition]{Definition--Theorem}
\newtheorem{definition-proposition}[proposition]{Definition--Proposition}

\theoremstyle{definition}

\theoremstyle{plain}

\theoremstyle{plain}

\numberwithin{equation}{section}
\numberwithin{proposition}{section}
\numberwithin{conj}{section}	

\renewcommand{\emptyset}{\ensuremath{\varnothing}}	

\usepackage{amsmath}
\usepackage[nottoc]{tocbibind}

\usepackage{geometry}
\usepackage{lmodern}

\title{Generalized Cardy conditions of topological defect lines}

\author[a]{Xia Gu\thanks{gux19@mails.tsinghua.edu.cn}}
\author[a]{Xianjin Xie}

\affil[a]{Yau Mathematical Sciences Center, Tsinghua University, Beijing 100084, China}

\date{}

\makeatletter
\renewcommand{\l@section}{\@dottedtocline{1}{1.5em}{2.0em}}
\renewcommand{\l@subsection}{\@dottedtocline{2}{4.0em}{3.0em}}
\makeatother

\begin{document}

\maketitle

\begin{abstract}
We propose a systematic procedure to work out systems of topological defect lines (TDLs) in minimal models. The only input of this method is the modular invariant partition function of the corresponding model. For diagonal and permutation diagonal models, we prove there is a bijection between simple TDLs and primary fields preserving fusion rules. For block-diagonal models, we derive simple TDLs in the $3$-state Potts model as an example and obtain results that align with those from $3$D topological field theory methods.
\end{abstract}  


\tableofcontents
\section{Introduction}
Topological defect lines (TDLs) in two dimensional conformal field theories (CFT) have been studied extensively in recent years\cite{Chang:2018iay,chang2023topological}. Roughly speaking, TDLs are 1-dimensional objects embedded in the space-time manifold. The correlation functions involving a TDL is invariant under continuous deformations of it. A primitive example of TDL is the spin-flip operation in the critical Ising model, which in turn is an element of the $\mathbb{Z}_2$ symmetry. Actually TDLs are one of many types of the generalised symmetries\cite{Gaiotto_2015} which can not be described by traditional notion of groups. In particular, a set of TDLs can include non-invertible lines whereas group elements always have an inverse. Kramers-Wannier (KW) duality in the critical Ising model is an example of such non-invertible lines. This line, denoted by $N$, follows a fusion rule $N\times N=1+\eta$, where $1$ represents the identity line and $\eta$ represents the $\Z_2$ symmetry line.

Much alike ordinary symmetries, TDLs can put a strong constraint on the dynamics of the underlying theory. For example, in the critical and tricritical Ising model, the existence of the $N$ line can be used to constrain the spectrum of IR TQFT\cite{Chang:2018iay}. There are many attempts to find KW-like defects in various situations, including those in higher dimensions\cite{Choi_2022,Kaidi_2022,chen2023symtfts}. Also, constructions of non-invertible symmetries in more realistic models have been done\cite{choi2022noninvertible,cordova2022noninvertible}. In these cases, non-invertible symmetries often show up when one tries to cancel an anomalous ordinary symmetry.

Restricting ourselves again to the $2$-dimensional conformal field theory, a natural question we can ask is: How to identify all the possible topological defect lines and their fusion rules? As an answer to this question, we would like to use the defect Cardy condition (along with other consistency arguments) to solve the TDLs in minimal models in this work. By "solving", we mean determining the action of all the simple TDLs on the Hilbert space, given only the spectrum and the partition function of the theory. Our method bears some resemblance to that in Petkova and Zuber\cite{Petkova:2000ip}, although there they only consider all the possible twisted partition functions. We will say more about the relation (and difference) of this work to \cite{Petkova:2000ip} in the end of section \ref{block diagonal}. 

We note that, for minimal models, TDLs can also be solved through an abstract and mathematically rigorous approach, developed by Fuchs et al. \cite{Fuchs:2001am,Fuchs:2002cm,Fuchs:2003id,Fuchs:2004dz,Fuchs:2004xi,Fjelstad:2005ua}. There, TDLs are formulated as bi-modules over two Frobenius algebras, denoted by $A$ and $B$. Given the chiral part information, namely the modular tensor category, the consistent set of TDLs can be worked out by classifying the $AB$ bi-modules. This formulation allows us to connect TDLs along with the underlying CFT to \(3\)D topological field theories (TFT), where the action of TDLs on the primary fields can be computed.  

In particular, the fusion categories of TDLs for non-diagonal ($D$-series and $E$-series) minimal models can be obtained from the corresponding diagonal ($A$-series) one through a functor. This functor maps a Frobenius algebra of the $A$-series category to the identity in the $D/E$-series category. In physics literature, this functor is referred to as the anyon condensation or the gauging of non-anomalous symmetries. Our treatment of non-diagonal models is independent from the gauging or anyon-condensation approach since it uses input only from the original non-diagonal theory. For the same reason, we can also employ this $3$D topological field theory(TFT) construction as a double-check that we have exhausted all the solutions.

The paper is organised as follows: In Section \ref{defct Cardy condition}, we summarize neccessary information about TDLs in \(2\)D CFT and the Cardy condition. Section \ref{diagonal and permutation model} and Section \ref{block diagonal} are dedicated to solving the defect Cardy condition in various models. Section \ref{TFT/CFT} introduces the TFT construction of Rational Conformal Field Theory and elucidate the results in the previous section from a different point of view. We recapitulate our results and comment on the possible applications in Section \ref{conclusion}.

\section{Defect Cardy condition}\label{defct Cardy condition}

In \(2\)D CFTs, the Hilbert space \(\mathcal{H}\) forms a representation of the Virasoro algebra, 
\begin{equation}
    [L_m,L_n]=(m-n)L_{m+n}+\frac{c}{12}(m^3-m)\delta_{m+n,0}
\end{equation}
where $L_m$ are the Fourier modes of the stress tensor \(T\). For simplicity, here we only talk about the holomorphic part of the theory. We will recover the anti-holomorphic story when it is necessary.  

Pick a time direction on the spacetime manifold, if a TDL is along the space direction, then it can be seen as a  linear operator on the Hilbert space:
\begin{align}
  \mathcal{L}:\mathcal{H}&\rightarrow\mathcal{H},\\
  \left|\phi\right\rangle&\mapsto \mathcal{L} \left|\phi\right\rangle.
\end{align}
The continuous deformations of TDL do not change the correlation function. This property, also referred to as isotopy invariance, implies that TDLs commute with all the Virasoro modes:
\begin{equation}
\label{eq:definition}
[\mathcal{L},L_m]=0,\;\forall m.
\end{equation}
The fusion of TDLs living in the same time-slice can be simply treated as multiplications of operators. That is, we can define the product line $\mathcal{L}\times \mathcal{L}'$ as 
\begin{equation}
    (\mathcal{L}\times \mathcal{L}')\ket{\phi}=\mathcal{L}\mathcal{L}'\ket{\phi}.
\end{equation}
Note that, this fusion is not necessarily commutative. The set of TDLs are closed under the fusion. In other words for any $\mathcal{L},\mathcal{L}'$, their product $\mathcal{L}\mathcal{L}'$ is also a TDL.

There is a set of simple TDLs $\mathcal{L}_{simple}$, whose non-negative-integer linear combinations can represent every TDL, while they can not represent each other. They play a role of ``basis'' in the ``vector spaces'' of TDLs. Note that, the fusion of two simple TDLs may not be simple, but can always be represented by non-positive-integer combinations. Finding the set of simple TDLs will be the goal of this work.  

Let $\ket{\phi}$ be a primary state, then from the isotopy invariance, we obtain the following formula:
\begin{equation}
\label{eq:eigen}
    \begin{aligned}
    L_m\mathcal{L}\ket{\phi}&=\mathcal{L}L_m\ket{\phi}=0,\;\forall m>0,\\
    L_0\mathcal{L}\ket{\phi}&=\mathcal{L}L_0\ket{\phi}=h_{\phi}\mathcal{L}\ket{\phi},
    \end{aligned}
\end{equation}
where \(h_{\phi}\) is the conformal weight of \(\ket{\phi}\). We find that $\mathcal{L}\ket{\phi}$ is also a primary state with the same conformal dimension as $\ket{\phi}$. So if there is no degeneracy about the conformal dimension $h$, the primary states $\ket{\phi}$ are eigenvectors of TDLs. Moreover, it can be easily shown that $\ket{\phi}$'s descendants are also eigenvectors of $\mathcal{L}$ of the same eigenvalue with $\ket{\phi}$ itself. 

Primaries and descendants form basis of the Hilbert space, in which the $\mathcal{L}$ can be represented as (infinite-rank)matrices. In the usual setting with no conformal dimension degeneracy, this matrix is diagonal and the essential information is contained in the entries corresponding to the primaries. We will discuss the case with conformal dimension degeneracy in Sec.\ref{block diagonal}.

Now consider a CFT defined on a torus. Define the twined partition function with a TDL $\mathcal{L}$ as:
\begin{equation}
\label{eq:generaltwined}
  Z^{\mathcal{L}}(\tau,\bar{\tau})=\Tr_{\mathcal{H}}{\mathcal{L} q^{L_0-\frac{c}{24}}\bar{q}^{\bar{L}_0-\frac{c}{24}}},\\
\end{equation}
where \(\tau\) is the modular parameter and \(q=e^{2\pi i \tau}\). This corresponds to a configuration in which $\mathcal{L}$ forming a spacial circle.

Then consider a modular $S$ transformation. The space and time direction are exchanged, so $\mathcal{L}$ is along the time circle and we get a twisted partition function $Z_{\mathcal{L}}$: 
\begin{equation}
\label{eq:modulartrans}
    Z_{\mathcal{L}}(\tau,\bar{\tau})=Z^{\mathcal{L}}(-\frac1\tau,-\frac{1}{\bar{\tau}}).
\end{equation}
$Z_{\mathcal{L}}$ is a partition function on the twisted Hilbert space $\mathcal{H}_{\mathcal{L}}$. In other words,
\begin{equation}
\label{eq:twistedPF}
  Z_{\mathcal{L}}(\tau,\bar{\tau})=\Tr_{\mathcal{H_L}}q^{L_0-\frac{c}{24}}\bar{q}^{\bar{L}_0-\frac{c}{24}}=M_{jk}\chi_{j}(\tau)\bar{\chi}_k(\bar{\tau}).\\
\end{equation}
where $\chi_j(\tau)$ and $\bar{\chi}_{k}(\bar{\tau})$ are Virasoro characters. $j$ and $k$ together label the primary states in the Hilbert space $\mathcal{H}_{\mathcal{L}}$. Elements of $M$ should be non-negative integers since in Hilbert space we have non-negative integer number of primary states:
\begin{equation}\label{eqNIMconstrain}
    M_{jk}\in \Z_{\geq 0}.
\end{equation}

\eqref{eq:modulartrans} along with \eqref{eqNIMconstrain} is our defect Cardy condition. This condition puts stringent restrictions on $\mathcal{L} $. We will use this condition to determine the possible $\mathcal{L}$ in various situations in the next section. 

\section{Diagonal and permutation-diagonal minimal models}\label{diagonal and permutation model}
Diagonal minimal models are the simplest cases which we study as a prototypical example. In such CFTs,  the partition functions are 
\begin{align}
    Z = \sum_{i} \chi_i(\tau) \bar\chi_{i}(\bar\tau).
\end{align}
The vacuum field is labeled by \(i=1\).
In diagonal theories, the action of $\mathcal{L}$ is completely determined by its eigenvalues of primaries, which we would solve. Below We will use $\mathcal{L}(i)$ to denote the eigenvalue of a general TDL $\mathcal{L}$ corresponding to the primary $\phi_i$.

There is a distinguished set of TDLs, called Verlinde lines, which is known to satisfy the defect Cardy condition. In diagonal minimal models, Verlinde lines are given by the familiar modular-S matrices\cite{Chang:2018iay}:
\begin{equation}
\label{eq:Verlindeline}
\mathcal{L}_m(i)=\frac{S_{im}}{S_{i1}},
\end{equation}
where $m$ is the label of Verlinde lines. Note that, the number of Verlinde lines is the same as that of primaries. More generally, the Verlinde lines are defined by the property of commuting with the maximal chiral algebra(which includes the Virasoro algebra) of the theory. We will deal with cases with extended chiral symmetries(block-diagonal models) in the latter section. 

Now we would like to prove this claim: In diagonal models, the Verlinde lines exhaust all the simple TDLs, and there exists a one-to-one map between Verlinde lines and primaries which preserves their fusion rules. In other words, every TDL can be written as non-positive-integer linear combinations of Verlinde lines. 
  
In diagonal models, the twined partition function \eqref{eq:generaltwined} reduces to
\begin{equation}
    Z^{\mathcal{L}}(\tau,\bar{\tau})=\sum_i\mathcal{L}(i)\chi_i(\tau)\bar{\chi}_i(\bar{\tau})
\end{equation}
From \eqref{eq:modulartrans}, we get
\begin{align}
\label{eq:origin}
M_{jk} = \sum_i \mathcal{L}(i) S_{i j}S^{*}_{ik},
\end{align}
which is equivalent to 
\begin{equation}
\label{eq:diagonal}
\mathcal{L}(i)\delta_{il}=\sum_{jk}S_{ij}^*M_{jk}S_{kl}
\end{equation}
We split \eqref{eq:diagonal} into
\begin{equation}
\label{eq:constraint}
\begin{gathered}
\sum_{j,k}S_{ij}^*M_{jk}S_{kl}=0\quad \text{for}\quad i\neq l,\\
    \mathcal{L}(i)=\sum_{j,k}S_{ij}^*M_{jk}S_{ki}.
\end{gathered}
\end{equation}
The first line of \eqref{eq:constraint} has the following solution
\begin{equation}
\label{eq:solution}
M_{jk}=M_{1m}N^k_{mj},
\end{equation}
where $N^k_{mj}$ is the corresponding fusion coefficients. This can be checked by inserting it back into the left hand side of \eqref{eq:constraint} and using the Verlinde formula $N_{mj}^k=\sum_n\frac{S_{nk}^*S_{nm}S_{nj}}{S_{n1}}$:
\begin{align}\label{solution checking}
\begin{split}
\sum_{j,k}S_{ij}^*(M_{1m}N_{mj}^k)S_{kl}&=M_{1m}S^*_{ij}S_{kl}\sum_n\frac{S_{nk}^*S_{nm}S_{nj}}{S_{n1}}\\
&=M_{1m}\sum_n\delta_{in}\delta_{nl}\frac{S_{nm}}{S_{n1}}\\
&=\delta_{il}M_{1m}\frac{S_{im}}{S_{i1}}.
\end{split}
\end{align}
Indeed, we can see that the last line is equal to zero unless $i=l$.
  
The Cardy condition requires $M_{jk}$ to be non-negative integers. Now this can be ensured as long as we let $M_{1m}$ to be non-negative integers since the fusion coefficients are also non-negative integers. Moreover, the whole solution sets of the Cardy condition are parametrized by $N$ non-negative integers $M_{1m},1\leq m\leq N$, where $N$ is the number of primary fields in the model. This is because we started with $\frac{N(N+1)}{2}$ unknowns $M_{jk}$, then solved $\frac{N(N-1)}{2}$ relations(the first line of \eqref{eq:constraint}), so the solution space is $N$ dimensional in $\R$. The $\R$-basis can be written as $(M_{11},M_{12},...,M_{jk},...)=(1,0,...,N_{1j}^k,...),(0,1,...,N_{2j}^k,...)$ and so on. Any solutions can be written as $\R$-combinations of them. Assume there is a solution $\vec{n}$ whose $m_{ij}$-entries are all positive integers, but its first coordinate $a_1$ in the above basis are not positive integers. Then $\Vec{n}=(a_1,...)$, obviously contradicting with the assumption. From this we can infer that all the coordinates of $\Vec{n}$ must be integers.

This means any solutions of the first lines of \eqref{eq:constraint} with non-negative integer entries $M_{jk}$ will be non-negative integer combinations of these basis. Since the second line of \eqref{eq:constraint} is linear in $M_{jk}$, every TDL can be written as non-integer linear combinations of solutions corresponding to these basis. In other words, these basis correspond to the set of simple TDLs. 

Now pick a solution in the basis: let $M_{1m}=1$ for a specific $m$, and $M_{1n}=0$ for $n\neq m$. Then 
\begin{equation}
\begin{aligned}
        \mathcal{L}(i)&=\sum_{j,k}S_{ij}^*N_{mj}^kS_{ki}\\
    &=\sum_{j,k,n}S^*_{ij}\frac{S_{nk}^*S_{nm}S_{nj}}{S_{n1}}S_{ki}\\
    &=\sum_n\delta_{in}\delta_{in}\frac{S_{nm}}{S_{n1}}\\
    &=\frac{S_{im}}{S_{i1}}.
\end{aligned}
\end{equation}
This is precisely the $m$th Verlinde line. 
We can conclude that the Verlinde lines are all the simple TDLs. By \eqref{eq:Verlindeline}, the label $m$ also corresponds to primary fields, so the number of simple TDLs is the same with that of the primaries. The fusion rules of Verlinde lines are:
\begin{equation}
\begin{aligned}
    \mathcal{L}_m(i)\times \mathcal{L}_n(i)&=\frac{S_{im}S_{in}}{S_{i1}S_{i1}} \\
    &=\sum_l \delta_{li}\frac{S_{lm}S_{ln}}{S_{l1}S_{i1}}\\
    &=\sum_{l,k}\frac{S^*_{lk}S_{lm}S_{ln}}{S_{l1}}\frac{S_{ik}}{S_{i1}}\\
    &=\sum_kN_{mn}^k \mathcal{L}_k(i),
\end{aligned}
\end{equation}
indeed same with that of primaries. 
\subsection{Permutation diagonal cases}
In this case, the modular invariant partition function is
\begin{align}
    Z = \sum_{i} \chi_{\sigma(i)}(\tau) \bar\chi_{i}(\bar\tau).
\end{align}
where $\sigma$ is an element of the permutation group of $S_n$, $n$ being the number of primaries. In this case the set of simple TDLs also has a fusion-rule preserving bijection with primary fields. The proof is a straightforward adaptation of the diagonal one so we will omit it. The only new ingredients here is the following property of modular matrices in the permutation diagonal models:
\begin{equation}
S_{ij}=S_{\sigma(i)\sigma(j)},
\end{equation}
and a delta function property $\delta_{ij}=\delta_{\sigma(i)\sigma(j)}$. The action of simple TDLs labeled by $m$ on the primary $\phi_i$, with a character $\chi_{\sigma(i)}\bar{\chi}_i$, can be eventually computed as: 
\begin{equation}
    \mathcal{L}_m(i)=\frac{S_{\sigma(i)m}}{S_{\sigma(i)0}}.
\end{equation}

\section{Block-diagonal minimal models}\label{block diagonal}
For general block-diagonal minimal models, there is no simple bijection rule between TDLs and primary fields. We would like to show this in the 3-state Potts model. But before this, let's do some preliminary analysis on partition functions in a general setting.

The block-diagonal minimal models can always be viewed as a diagonal one under an extended chiral algebra $\mathcal{A}$ by combining Virasoro characters together.
This is to say, we have partition functions in the following form:
\begin{equation}
    Z=\sum_\alpha B_\alpha(\tau) \bar{B}_\alpha(\bar{\tau}),
\end{equation}
where $\alpha$ is the label of $\mathcal{A}$-primaries and $B_\alpha$ are $\mathcal{A}$ characters. And of course, $B_\alpha$ can be decomposed into Virasoro characters:
\begin{equation}
    B_\alpha=C_{\alpha l}\chi_l.
\end{equation}
where $C_{\alpha l}$ is the combination coefficients. So the partition function can also be written as 
\begin{equation}
 Z=\sum_{\alpha,l,k}C_{\alpha l}C_{\alpha k}\chi_l\bar{\chi}_k.
\end{equation}

As characters, $B_\alpha$ transform among themselves under the modular group and we will use $S^{ex}$ to denote the corresponding modular matrices. Consider the modoular transformation:
\begin{equation}
    \begin{aligned}
        Z&=S^{ex}_{\alpha\beta}\bar{S}^{ex}_{\alpha\gamma}B_{\beta}\bar{B}_{\gamma}\\
        &=S^{ex}_{\alpha\beta}\bar{S}^{ex}_{\alpha\gamma}C_{\beta l}C_{\gamma k}\chi_{l}\bar{\chi}_{k}\\
        &=C_{\alpha i}C_{\alpha j}S_{il}\bar{S}_{jk}\chi_l\bar{\chi}_k.
    \end{aligned}
\end{equation}
This leads to a relation of coefficients $C_{\alpha l}$ and modular matrices $S$:
\begin{equation}
\label{eq:modular}
(S^{ex})_{\alpha\beta}C_{\beta l}=C_{\alpha i}S_{il}
\end{equation}
along with its complex conjugation. In principle, knowing $C_{\alpha m}$ and $S_{ml}$, we can compute $S^{ex}$ accordingly.

Consider lines(operators) given by this extended modular matrix:
\begin{equation}
   \label{eq:Verlinde} \mathcal{L}_{\alpha}\ket{\varphi_\beta}=\frac{(S^{ex})_{\beta\alpha}}{(S^{ex})_{\beta 0}}\ket{\varphi_\beta},
\end{equation}
where $\varphi_\beta$ denotes $\mathcal{A}$-primaries. Now the conclusion of the last section can be directly applied here. That is, the lines defined by \eqref{eq:Verlinde} exhaust all the Verlinde lines, or lines that commute with the maximal chiral algebra $\mathcal{A}$. 

The Verlinde lines will be our starting point of computing all the other TDLs. It turns out that to completely determine the TDLs in the block-diagonal model, one should consider not only the Cardy condition but also fusion rules.


\subsection{TDLs in \texorpdfstring{$3$}{3}-State Potts Model}
$3$-State Potts model can be seen as a diagonal model with extended $W_3$ symmetry. There are $6$ $W_3$ primary fields $1,\epsilon, Z, Z^*, \sigma, \sigma^*$.  Accordingly, there should be $6$ Verlinde lines. But when using \eqref{eq:Verlinde} one need to be careful that $Z$ and $Z^*$, also $\sigma$ and $\sigma^*$ have the conformal dimension, hence the same character. We will see that this degeneracy is the chief reason that complicates the analysis of the Cardy condition.

Due to this degeneracy, if we want to obtain a $6\times 6$ modular matrix $S^{W_3}$, we should first use \eqref{eq:modular} to get a $4\times 4$ matrix and split it carefully, preserving the symmetry and unitarity of the modular matrix. The explicit form of this $S^{W_3}$ matrix can be found in chapter 10 of \cite{DiFrancesco:1997nk}, which we copy below for reference:
\begin{equation}
\begin{aligned}
  S^{W^3}=
    \frac{2}{\sqrt{15}}\begin{pmatrix}
      s_1&s_2&s_1&s_1&s_2&s_2\\
      s_2&-s_1&s_2&s_2&-s_1&-s_1\\
      s_1&s_2&\omega s_1&\omega^2s_1&\omega s_2&\omega^2 s_2\\
      s_1&s_2&\omega^2 s_1&\omega s_1&\omega^2 s_2&\omega s_2\\
      s_2&-s_1&\omega s_2&\omega^2 s_2&-\omega s_1&-\omega^2 s_1\\
      s_2&-s_1&\omega^2 s_2&\omega s_2&-\omega^2 s_1&-\omega s_1
    \end{pmatrix}   \\
s_1=\sin(\pi/5),\;s_2=\sin(2\pi/5),\;\omega=e^{2i\pi/3}.
\end{aligned}
\end{equation}
Now the Verlinde lines can be computed by \eqref{eq:Verlinde}. The eigenvalue of each line with respect to 6 primaries
$1,\epsilon,Z,Z^*,\sigma,\sigma^*$ are shown in Table.\ref{tab:W3Verlinde}.

\begin{table}
    \centering
    \begin{tabular}{c|c|c|c|c|c|c}
Lines&1&$\epsilon$&$Z$&$Z^*$&$\sigma$&$\sigma^*$\\
\hline
   1&1&1&1&1&1&1\\       $\eta$&1&1&$\omega$&$\omega^2$&$\omega$&$\omega^2$\\
 $\eta^2$&$1$&$1$&$\omega^2$&$\omega$&$\omega^2$&$\omega$\\
 $W$&$\zeta$&$-\zeta^{-1}$&$\zeta$&$\zeta$&$-\zeta^{-1}$&$-\zeta^{-1}$\\
 $W\eta$&$\zeta$&$-\zeta^{-1}$&$\omega\zeta$&$\omega^2\zeta$&$-\omega\zeta^{-1}$&$-\omega^2\zeta^{-1}$\\
 $W\eta^2$&$\zeta$&$-\zeta^{-1}$&$\omega^2\zeta$&$\omega\zeta$&$-\omega^2\zeta^{-1}$&$\omega\zeta^{-1}$
    \end{tabular}
    \caption{Verlinde lines in the three-state Potts model. Here $\zeta=\frac{\sqrt{5}+1}{2}$.}
    \label{tab:W3Verlinde}
\end{table}
    
Their fusion rules are the same with the fusion rules of $W_3$ primaries, which can be directly verified by multiplying eigenvalues in Table.\ref{tab:W3Verlinde}.

Now we turn to general simple TDLs which we would like to solve using the Cardy condition.
Unlike in the diagonal model case, where there is no degeneracy, the action of $\mathcal{L}$ on primaries $\phi_i$ may not be diagonal. Recalling \eqref{eq:eigen}, all we can say now is $\mathcal{L}\ket{\phi}$ has the same $h$ with $\ket{\phi}$. It may mix different primary states with the same $h$. 

In 3-state Potts model, there are $12$ Virasoro primaries, including $Z,Z^*$ and $\sigma,\sigma^*$. Other fields all have distinct conformal dimensions $(h,\bar{h})$. The action of $\mathcal{L}$ may mix $Z$ with $Z^*$, also $\sigma$ with $\sigma^*$, but no other states. So we need $8$ eigenvalues, corresponding to other $8$ primaries, and two rank-$2$ block matrices to characterize the matrix $\mathcal{L}$.

We will use notations in \eqref{eq:families} to represent the whole matrix. Here the first parenthesis encloses those $8$ eigenvalues, and in the following 2 blocks contain off-diagonal entries mixing $Z$ and $Z^*$($\sigma$ and $\sigma^*$).    
\begin{equation}
\label{eq:families}
    \mathcal{L}\doteq (......)\begin{pmatrix}
        \mathcal{L}_{ZZ}& \mathcal{L}_{ZZ^*}\\
        \mathcal{L}_{Z^*Z}&\mathcal{L}_{Z^*Z^*}
    \end{pmatrix}
    \begin{pmatrix}
        \mathcal{L}_{\sigma\sigma}& \mathcal{L}_{\sigma\sigma^*}\\
        \mathcal{L}_{\sigma^*\sigma}&\mathcal{L}_{\sigma^*\sigma^*}.
    \end{pmatrix}
\end{equation}

The twined partition function is as follows:
\begin{equation}
\begin{aligned}
Z^{\mathcal{L}}&=\sum_{p,dec}\bra{\phi_p}q^{L_0-\frac{c}{24}}\bar{q}^{\bar{L}_0-\frac{\bar{c}}{24}}\mathcal{L}\ket{\phi_p}\\
&=\sum_{p,r,dec}\bra{\phi_p}q^{L_0-\frac{c}{24}}\bar{q}^{\bar{L_0}-\frac{\bar{c}}{24}}\mathcal{L}_{pr}\ket{\phi_r}\\
&=\sum_p \mathcal{L}_{pp}\chi_p\bar{\chi}_p+(\mathcal{L}_{ZZ}+\mathcal{L}_{Z^*Z^*})\chi_{Z(Z^*)}\bar{\chi}_{Z(Z^*)}+(\mathcal{L}_{\sigma\sigma}+\mathcal{L}_{\sigma^*\sigma^*})\chi_{\sigma(\sigma^*)}\bar{\chi}_{\sigma(\sigma^*)}.
    \end{aligned}
\end{equation}
The notation here needs some explanation. We use $dec$ to denote the summation over descendants. $p$ and $r$ label primaries. In block-diagonal models, the correspondence between primaries and characters is not one-to-one. By $\chi_p$ we mean the holomorphic character of the $p$th primary, so there may be another primary which has the same character. Also, in this notation $\chi_p$ and $\bar{\chi}_p$ is not necessarily a conjugate pair. In the last line, the summation label $p$ only runs from $1$ to $8$, corresponding to the first $8$ eigenvalues of $\mathcal{L}$.    

This already tells us that the Cardy condition alone cannot fully determine the $\mathcal{L}$ matrix, since in the partition function, off-diagonal entries' contribution is missing. Also, the coefficient before $\chi_{Z(Z^*)}\bar{\chi}_{Z(Z^*)}$($\chi_{\sigma(\sigma^*)}\bar{\chi}_{\sigma(\sigma^*)}$) is $\mathcal{L}_{ZZ}+\mathcal{L}_{Z^*Z^*}$($\mathcal{L}_{\sigma\sigma}+\mathcal{L}_{\sigma^*\sigma^*}$) bound together, can not be solved independently. 

The twisted partition function can still be written as 
\begin{equation}
Z_{\mathcal{L}}=M_{jk}\chi_j\bar{\chi}_k.
\label{defectCardyIII}
\end{equation}
where $j$ and $k$ label characters, different from $p$. 
There is a set of equations of $M_{jk}$ and $\mathcal{L}_{pp}$ related by modular matrices analogous to \eqref{eq:constraint}. Using the first line, we can reduce the number of independent integers $M_{jk}$ to $10$ \cite{haghighat2023topological}. Unfortunately, there is no convenient expressions of $M_{jk}$ in block-diagonal model as in \eqref{eq:solution} though we can choose some elementary $M_{jk}$s and represent others by a non-negative integer linear combination. We will put the details about $M_{jk}$ in the appendix.

The next step is to use the analogous version of the second line of \eqref{eq:diagonal} to compute $\mathcal{L}_{pr}$. Although this procedure is straightforward, due to the caveat mentioned before, we can only restrict the solutions of $\mathcal{L}$ to $10$ continuous families with undetermined parameters, corresponding to $10$ chosen non-negative integers among $M_{lk}$. For simplicity we will denote the value of two sums $\mathcal{L}_{ZZ}+\mathcal{L}_{Z^*Z^*}$, $\mathcal{L}_{\sigma\sigma}+\mathcal{L}_{\sigma^*\sigma^*}$ as $\text{tr}_Z$ and $\text{tr}_\sigma$, respectively. And the $10$ families are(invoking the notation in \eqref{eq:families}):
\begin{equation}
\label{eq:10solu}
\begin{aligned}
   [1]&:(1,1,1,1,1,1,1,1)\quad \Tr_Z=\Tr_\sigma=2\\
[\eta]&:(1,1,1,1,1,1,1,1)\quad \Tr_Z=\Tr_\sigma=-1\\
   [W]&:(\zeta,\zeta,\zeta,\zeta,-\zeta^{-1},-\zeta^{-1},-\zeta^{-1},-\zeta^{-1})\quad \Tr_Z=2\zeta\quad\Tr_\sigma=-2\zeta^{-1} \\           
[W\eta]&:(\zeta,\zeta,\zeta,\zeta,-\zeta^{-1},-\zeta^{-1},-\zeta^{-1},-\zeta^{-1})\quad \Tr_Z=-\zeta\quad \Tr_\sigma=-\zeta^{-1}\\
[C]&:(1,1,-1,-1,1,1,-1,-1)\quad \Tr_Z=\Tr_\sigma=0\\
    [CW]&:(\zeta,-\zeta,-\zeta,\zeta,-\zeta^{-1},\zeta^{-1},-\zeta^{-1},\zeta^{-1})\quad \Tr_Z=\Tr_\sigma=0\\
    [N]&:(\sqrt{3},-\sqrt{3},\sqrt{3},-\sqrt{3},-\sqrt{3},\sqrt{3},-\sqrt{3},\sqrt{3})\quad \Tr_Z=\Tr_\sigma=0\\
    [WN]&:(\sqrt{3}\zeta,-\sqrt{3}\zeta,\sqrt{3}\zeta,-\sqrt{3}\zeta,\sqrt{3}\zeta^{-1},-\sqrt{3}\zeta^{-1},\sqrt{3}\zeta^{-1},-\sqrt{3}\zeta^{-1})\quad \Tr_Z=\Tr_\sigma=0\\
    [CN]&:(\sqrt{3},-\sqrt{3},-\sqrt{3},\sqrt{3},-\sqrt{3},\sqrt{3},\sqrt{3},-\sqrt{3})\quad \Tr_Z=\Tr_\sigma=0\\
    [CWN]&:(\sqrt{3}\zeta,-\sqrt{3}\zeta,-\sqrt{3}\zeta,\sqrt{3}\zeta,\sqrt{3}\zeta^{-1},-\sqrt{3}\zeta^{-1},-\sqrt{3}\zeta^{-1},\sqrt{3}\zeta^{-1})\quad \Tr_Z=\Tr_\sigma=0.
\end{aligned}
\end{equation}
The square brackets with symbols, for example $[1]$, denote the whole continuous family. The meaning of the symbols will be clear in the end of this section. Note that, $\mathcal{L}_{ZZ^*},\mathcal{L}_{Z^*Z}$ and $\mathcal{L}_{\sigma\sigma^*},\mathcal{L}_{\sigma^*\sigma}$ are left completely free in these families. 

Following the similar argument with the diagonal case, we know any simple TDL must fall into one of the families. That is, it must have one set of the given first $8$ eigenvalues and given $\tr_Z,\tr_\sigma$. 

Recall that, for any two simple TDL $\mathcal{L},\mathcal{L}'$, their product is also a TDL, hence can be decomposed as simple TDLs. 
This can provide new restrictions on parameters $\mathcal{L}_{ZZ^*}$ and so on of simple TDLs, since we need to impose the condition of $\Tr_Z$ and $\Tr_\sigma$ again on the resulted new lines. By repeatedly fusing simple lines, decomposing the results into new lines and considering consistency of fusion rules, we can eventually compute the parameters and prove these families are all finite set.

We will focus on solving the entries $\mathcal{L}_{ZZ},\mathcal{L}_{ZZ^*},\mathcal{L}_{Z^*Z}$ and $\mathcal{L}_{Z^*Z^*}$, referring to $\begin{pmatrix}
    \mathcal{L}_{ZZ}& \mathcal{L}_{ZZ^*}\\
        \mathcal{L}_{Z^*Z}&\mathcal{L}_{Z^*Z^*}
\end{pmatrix}$ as $\mathcal{L}_{Zblock}$. The $\sigma$-entries can be computed independently in a similar way so we omit the details. The analysis in the rest of this section is very dense.    

Let's start from the family $[1]$. Consider a simple line in $[1]$, and call it $I$. Consider $I\times I$, of which the first $8$ diagonal elements are all $1$. So $I\times I$ must be a simple line either in $[1]$ or $[\eta]$. 

Assume $I\times I\in[\eta]$, then from the trace condition $\Tr_Z(I\times I)=-1$, which we can solve to express $I_{ZZ}$ and $I_{Z^*Z^*}$ in terms of $I_{ZZ^*}$ and $I_{Z^*Z}$. The solution is:
\begin{equation}
    I_{Z block}=\begin{pmatrix}
        1+\sqrt{-\frac32+I_{ZZ^*}I_{Z^*Z}}&I_{ZZ^*}\\
        I_{Z^*Z}&1-\sqrt{-\frac{3}{2}+I_{ZZ^*}I_{Z^*Z}}
    \end{pmatrix}.
\end{equation}
In this case the trace of $I^3_{Zblock}$ is $-7$, neither $2$ nor $-1$, though their first $8$ diagonal elements are all $1$. This indicates $I^3$ does not fall in these families, which is an inconsistency. This excludes the possibility of $I\times I\in[\eta]$.

So $I\times I\in [1]$. Solving $\tr_Z=2$, we have 
\begin{equation}
    I_{Z block}=\begin{pmatrix}
        1+i\sqrt{I_{ZZ^*}I_{Z^*Z}}&I_{ZZ^*}\\
        I_{Z^*Z}&1-i\sqrt{I_{ZZ^*}I_{Z^*Z}}
    \end{pmatrix}.
\end{equation}
Moreover, 
\begin{equation}
(I^k)_{Zblock}=\begin{pmatrix}
        1+k i\sqrt{I_{ZZ^*}I_{Z^*Z}}&kI_{ZZ^*}\\
        kI_{Z^*Z}&1-k i\sqrt{I_{ZZ^*}I_{Z^*Z}}
    \end{pmatrix},
\end{equation}
hence $I^k\in[1]$ for all $k$.
With a little bit of reasoning, we can also infer that if $I^k=1$ for some $k$, then $I=1$.

From now on we will speed up and sort lines into families without mentioning the words ``comparing the first $8$ diagonal elements''. We will also drop ``from the trace condition'' unless we want to stress. We would also like to use the Verlinde lines as the known solution. 

$I\times \eta$ can either be in $[1]$ or $[\eta]$. If it is in $[1]$, then from the trace condition $\sqrt{I_{ZZ^*}I_{Z^*Z}}=\frac{3}{i(\omega-\omega^2)}$. In this case $(I\times \eta^2)_{Zblock}$ have trace $-4$, meaning the assumption is false. 

So $I\times \eta\in [\eta]$, this gives $\sqrt{I_{ZZ^*}I_{Z^*Z}}=0$. Without loss of generality, we can assume $I_{Z^*Z}=0$. So now
\begin{equation}
\label{eq:1line}
I_{Zblock}=\begin{pmatrix}
        1&I_{ZZ^*}\\
        0&1
\end{pmatrix}.
\end{equation}

If there is a line $I'\in[1]$ which takes the form
\begin{equation}
    I'_{Zblock}=\begin{pmatrix}
        1&0\\
        I'_{Z^*Z}&1
    \end{pmatrix},
\end{equation}
consider $I\times I'$, of which the trace is $I_{ZZ^*}I'_{Z^*Z}+2$. Suppose $I\times I'\in[\eta]$, then $I_{ZZ^*}I'_{Z^*Z}=-3$. In this case $\Tr_Z(I^2\times I')=-4$, meaning the assumption is false.

Suppose $I\times I'\in[1]$ Then $I_{ZZ^*}I'_{Z^*Z}=0$. Our assumption is $I_{ZZ^*}\neq0$ in general. So $I'_{Z^*Z}=0$. We can conclude that, simple lines in $[1]$ are all in the form of \eqref{eq:1line}, and their fusion all lies in $[1]$.

Now we turn to $[\eta]$. We label a candidate simple line in $[\eta]$ as $E$. $E\times E\in[\eta]$ or $E\times E \in [1]$. In the latter case $\Tr_Z(E^3)=-\frac52$, which doesn't belong to any families.

So $E\times E\in [\eta]$. From the trace condition we get
\begin{equation}
\label{eq:etama}
\begin{gathered}
E_{Zblock}=\begin{pmatrix}
        \frac12(-1-\sqrt{-3-4E_{ZZ*}E_{Z^*Z}})& E_{ZZ^*}\\
     E_{Z^*Z}& \frac12(-1+\sqrt{-3-4E_{ZZ*}E_{Z^*Z}})
    \end{pmatrix},\\
    (E^2)_{Zblock}=\begin{pmatrix}
        \frac12(-1+\sqrt{-3-4E_{ZZ*}E_{Z^*Z}})& -E_{ZZ*}\\
        -E_{Z^*Z}& \frac12(-1-\sqrt{-3-4E_{ZZ*}E_{Z^*Z}})
    \end{pmatrix},\\ E^3=1.
\end{gathered}
\end{equation}
Note that, if $E_{ZZ^*}=E_{Z^*Z}=0$, $E$ reduces to $\eta$ or $
\eta^2$.

Now name a candidate simple in $[C]$ as $\mathcal{C}$.
Consider $I\times \mathcal{C}$ and $\mathcal{C}\times I$. These two lines lie in $[C]$. The trace condition gives 
\begin{equation}
\label{eq:ac}  I_{ZZ^*}\mathcal{C}_{ZZ^*}=I_{ZZ^*}\mathcal{C}_{Z^*Z}=0.
\end{equation}
 Then we fuse $\mathcal{C}$ with itself: 
\begin{equation}
    \mathcal{C}\times \mathcal{C}=(1,1,1,1,1,1,1,1)\quad \begin{pmatrix}                       \mathcal{C}_Z^2+\mathcal{C}_{Z^*Z}\mathcal{C}_{ZZ^*}&0\\
    0& \mathcal{C}_Z^2+\mathcal{C}_{Z^*Z}\mathcal{C}_{ZZ^*}
    \end{pmatrix}.
\end{equation}
We can see that $\mathcal{C}\times \mathcal{C}\in[1]$ or $\mathcal{C}\times \mathcal{C} \in [\eta]$. The latter case is not possible because we have determined the form of lines in $[\eta]$ as \eqref{eq:etama}. The 2 diagonal elements cannot be equal. 

So $\mathcal{C}\times \mathcal{C}\in[1]$. Recall \eqref{eq:1line}, we know $\mathcal{C}\times \mathcal{C}=1$. 
So
\begin{equation}
\label{eq:CC condition} \mathcal{C}_Z^2+\mathcal{C}_{Z^*Z}\mathcal{C}_{ZZ^*}=1.
\end{equation}

Now fuse this $\mathcal{C}$ with $\eta$, $\mathcal{C}\times \eta\in [C]$. This gives the constraint $\omega \mathcal
{C}_Z=\omega^2 \mathcal{C}_Z$ which leads to $\mathcal{C}_Z=0$. Using \eqref{eq:CC condition}, $\mathcal{C}_{Z^*Z}\mathcal{C}_{ZZ^*}=1$. 
Looking back to \eqref{eq:ac}, we conclude that $I_{ZZ^*}=0$, in other words $[1]=\{1\}$.

$\mathcal{C}\times \eta\neq \mathcal{C}$, otherwise $\eta=1$ from $\mathcal{C}\times \mathcal{C}=1$. For the same reason, $\mathcal{C}\times \eta^2$ is different from $\mathcal{C}$ and $\mathcal{C}\eta$. Using the matrix multiplication law, we can verify that $\eta \mathcal{C}=\mathcal{C}\eta^2$ and $\eta^2\mathcal{C}=\mathcal{C}\eta$. So lines in $[C]$ appear in triplets of $\mathcal{C},\mathcal{C}\eta,\mathcal{C}\eta^2$.

Now arbitrarily pick two lines from $[C]$, fuse them together:
\begin{equation}
\mathcal{C}\times \mathcal{C}'=(1,...,1)\begin{pmatrix}
    \mathcal{C}_{ZZ^*}^{-1}\mathcal{C}'_{ZZ^*}&0\\
    0&\mathcal{C}_{ZZ^*}\mathcal{C}_{ZZ^*}^{-1}
\end{pmatrix} =1 \; or \; \eta \; or \; \eta^2.
\end{equation}
So $\mathcal{C}_{ZZ^*}=\mathcal{C}'_{ZZ^*}$ or $\mathcal{C}_{ZZ^*}=\omega \mathcal{C}'_{ZZ^*}$ or $\mathcal{C}_{ZZ^*}=\omega^2\mathcal{C}'_{ZZ^*}$. This means they are from the same triplet. In a word, $[C]$ has only 3 elements $\{\mathcal{C},\mathcal{C}\eta,\mathcal{C}\eta^2\}$.

Actually, $\mathcal{C}_{Z^*Z}$ depends on the relative normalization of $\ket{Z}$ and $\ket{Z^*}$, so they can be freely chosen. We can safely adopt the traditional convention here, letting $[C]=\{C,C\eta,C\eta^2\}$ with $C_{ZZ^*}=1,\;C\eta_{ZZ^*}=\omega^2,\;C\eta^2_{ZZ^*}=\omega$.

Consider $C\times E\in [C]$. Now $[C]$ has only $3$ elements, so comparing the matrix gives us $E=\eta, E^2=\eta^2$. In other words $[\eta]=\{\eta,\eta^2\}$.

Having fixed $[1],[\eta],[C]$, lines in other families are relatively easy to determine. Consider a simple line candidate $\mathcal{W}$ in $[W]$, $C\times \mathcal{W}\in[CW]$ gives $\mathcal{W}_{ZZ^*}=-\mathcal{W}_{Z^*Z}$. Further, $C\times \mathcal{W}\times \eta\in[CW]$ gives $\mathcal{W}_{ZZ^*}=\mathcal{W}_{Z^*Z}=0$. Then $\mathcal{
W}\times \eta\in [W]\; or\;[W\eta]$ and $\mathcal{W}\times \eta^2\in [W]\;or\;[W\eta]$ gives $\mathcal{W}_{ZZ}=\zeta$. So there is only one line in $[W]$. Name it as $W$ again. 

Name a simple line candidate in $[W\eta]$ as $V$, the trace condition of $C\times V\in[CW]$ gives $V_{ZZ^*}=V_{Z^*Z}$. Further, $C\times V \times \eta\in[CW]\; or\;[C]$ gives $V_{ZZ^*}=V_{Z^*Z}=0$. Trace condition of $V\times \eta\in [W]\; or\;[W\eta]$ and $V\times \eta^2\in [W]\;or\;[W\eta]$ gives $V_{ZZ}=\omega\zeta\;or\;\omega^2\zeta$. So $[W\eta]=\{W\eta,W\eta^2\}$.

Consider a simple line $X$ in $[CW]$. $X\times \eta\in[CW]$ gives $X_{ZZ}=X_{Z^*Z^*}=0$. Consider simultaneously $C\times X,\;C\times X\times \eta$ and $C\times X\times \eta^2$. These lines fall into $[W]$ or $[W\eta]$. There are only $3$ consistent solutions: 
\begin{equation}
    \begin{aligned}
1.&X_{ZZ^*}=X_{Z^*Z}=\zeta, \\2.&X_{ZZ^*}=\omega^2\zeta,\;X_{Z^*Z}=\omega\zeta,\\ 3.&X_{ZZ^*}=\omega\zeta,\;X_{Z^*Z}=\omega^2\zeta.
    \end{aligned}
\end{equation}
 So we get $[CW]=\{CW,CW\eta,CW\eta^2\}$.

Now we turn to lines in $[N]$. Name a simple line candidate as $\mathcal{N}$. Consider $\mathcal{N}\times \eta\in[N]$. This gives the constraint $\omega \mathcal{N}_{ZZ}=\omega^2 \mathcal{N}_{ZZ}$, which means $\mathcal{N}_{ZZ}=0$. 

Then we fuse $C$ with $\mathcal{N}$. $C\mathcal{N}\in[CN]$, which leads to the constraint $\mathcal{N}_{Z^*Z}=-\mathcal{N}_{ZZ^*}$. Consider then $C\mathcal{N}\times \eta$. This line is also in $[CN]$, which gives $\omega  \mathcal{N}_{ZZ^*}=\omega^2  \mathcal{N}_{ZZ^*}$. So we can conclude $\mathcal{N}_{ZZ^*}=\mathcal{N}_{Z^*Z}=0$. There is only one line in $[N]$ with $Z$ block being a zero matrix. We call it $N$.

There are $3$ families $[WN],[CN],[CWN]$ left undetermined. But they can be done following exactly the same path of determining $[N]$. They all contain one line with $Z$ block being zero.  

At this point, we have found the same set of simple TDLs and fusion rules with those in \cite{haghighat2023topological,Chang:2018iay}. Moreover we have proved that these are the complete set. For the reference, we present the fusion rules here. The fusion ring is generated by $1,\eta,W,C,N$:
\begin{equation}
\begin{aligned}
      &1\times \mathcal{L}=\mathcal{L} \quad(\text{for any}\; \mathcal{L}),  \\
      &\eta^3=1,\\
      &W\times W=1+W,\\
      &C\times C=1,\\
      &C\times \eta=\eta^2\times C,\; C\times \eta^2=\eta\times C\quad(\text{non commutative}),\\
      &N\times N=1+\eta+\eta^2,\\
      &N\times \eta=\eta.
\end{aligned}
\label{fusionTP}
\end{equation}
\subsection{On general block-diagonal models}
We have seen from the example that, the ambiguous parameters in the TDLs before considering fusion rules are from fields with degeneracy. Suppose there are $m$ non-degenerate fields and $n$ set of degenerate fields in a general block-diagonal model. For each solution family, now the number of ambiguous parameter is roughly proportional to $n$. On the other hand, one fusion product will give exactly $n$ trace condition and every solution has $m+n$ families to fuse with. In total there will be $n(m+n)$ constraints for each solution families, which reasonably outnumber $~n$. We can count multiple fusions in, if fusing only once is not enough. This means the procedure can be carried out basically.

Also, some physics-related observations can help to simplify the procedure. In $3$ state Potts model, there are doubly-degenerate fields $Z,Z^*$ and $\sigma,\sigma^*$. Accordingly, the lines of charge conjugation $[C]$ arises, which permutes these fields. So in general models with degeneracy, we can expect there will be equivalences of $C$. Depending on the multiplicity of the degeneracy, $C$ may permute the fields in a more involved way. But once one can verify $C$ line do satisfy the Cardy condition, it will significantly promotes the procedure. 

Moreover, there is a $N$ duality line in the $3$ State Potts model, which fuses itself to the sum of $\mathbb{Z}_3$ symmetry lines. $N$ together with $1,\eta,\eta^2$ forms the Tambara-Yamagami category. From Verlinde lines computation, we can readily see if there is $\mathbb{Z}_n$ symmetry in the model. Models with degeneracy are very likely to have this symmetry since $\mathbb{Z}_n$ interacts with charge conjugation $C$ intimately. In this case, the $N$ line fusing itself to the sum of all the elements of $\mathbb{Z}_n$ probably exists, which in turn simplifies the counting of simple TDLs a lot. 

 Finally, we would like to comment on \cite{Petkova:2000ip}. There, the authors start from an assumption that the total number of simple TDLs are a finite number $n$, which is the sum of squares of multiplicities of the characters in the partition function. Further, They also assume that the $n\times n$ matrices whose rows are action values of individual simple TDLs are unitary. Then they derived a set of relations satisfied by twisted partition functions with 2 fused simple TDLs inserted. These twisted partition function are classified but no individual lines are worked out. Our approach don't assume a priori the number of simple TDLs, nor the eigenvalues of TDL-actions. And we have gone a step further to determine all the individual simple TDL actions.  

\section{TFT construction of RCFT}\label{TFT/CFT}

In this section, we utilize the topological field theory (TFT) approach, developed by Fuchs et al\cite{Fuchs:2001am, Fuchs:2002cm, Fuchs:2003id, Fuchs:2004dz, Fuchs:2004xi}, to validate our findings. In this approach, all the quantities in RCFTs corresponds to algebraic structures. We provide a summary of this algebraic aspects in the Appendix \ref{Algebraic aspects of RCFT and TDL}. In the following, we will use the terminology defined in the Appendix directly. 

\subsection{General formulation}
A \(3\)D TFT of Reshetikhin-Turaev type can be constructed from an modular tensor category (MTC)\cite{RT1990}. We are interested in the MTC \(\mathcal{C}\) of representations of an chiral algebra $\mathcal{V}$, namely, \(\mathcal{C}=\mathcal{R}ep(\mathcal{V})\). Inspired from Chern-Simons TFTs/Wess-Zumino-Witten models correspondence and the holomorphic factorization, we further expect that the correlators of RCFTs defined on a surface \(\Sigma\) are equivalent to the conformal blocks on its double.

Consider an orientable world sheet \(\Sigma\). We can double the manifold except the boundary and denoted it as \(\widehat{\Sigma}\). This mimics the mirror charge method in classical electrodynamics. For example, if \(\Sigma\) is a disc, \(\widehat{\Sigma}\) is a sphere. The space of conformal blocks for \(\widehat{\Sigma}\) with field insertions is given by evaluating the TFT functor $\text{RT}_\mathcal{C}$ on \(\widehat{\Sigma}\) with Wilson lines. In particular, the correlator \(\text{Cor}_\mathcal{C}(\Sigma)\) is obtained by evaluating \(\text{RT}_\mathcal{C}\) on a specific three-dimensional bordism \(M_\Sigma:  \emptyset \rightarrow \widehat{\Sigma}\) with an embedded \(\mathcal{C}\)-colored ribbon link (Wilson lines),
\begin{align}
\text{RT}_{\mathcal{C}}(M_\Sigma,\emptyset,\partial M_{\Sigma}):\text{RT}_{\mathcal{C}}(\emptyset)=\CC \rightarrow \text{RT}_{\mathcal{C}}(\widehat{\Sigma}),\quad 1\mapsto \text{Cor}(\Sigma) 
\end{align}

\noindent For our purposes, we need the following \(3\)D bordism:
\begin{align}
M_\Sigma\coloneqq \Sigma\times [-1,1]/\sim
\end{align}
where the equivalence relation \(\sim\) is given by \((x,t) \sim (x,-t)\) for all \(x\in \partial \Sigma,t\in[-1,1]\). 
Note that there is a natural embedding 
\begin{align*}
\iota:\Sigma&\hookrightarrow M_\Sigma\\
x&\mapsto (x,0).
\end{align*}

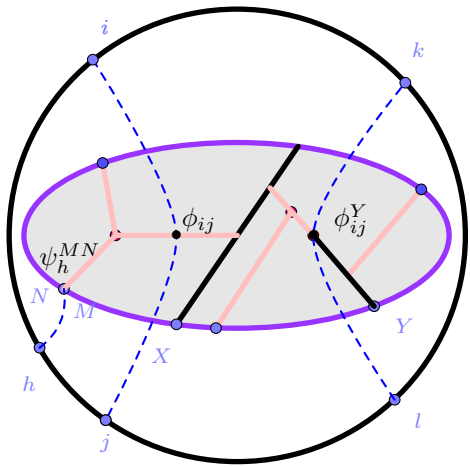
\begin{figure}[h!t]
\centering
\definecolor{ccqqqq}{rgb}{0.8,0,0}
\definecolor{xdxdff}{rgb}{0.49019607843137253,0.49019607843137253,1}
\definecolor{zzttff}{rgb}{0.6,0.2,1}
\definecolor{ududff}{rgb}{0.30196078431372547,0.30196078431372547,1}
\definecolor{uuuuuu}{rgb}{0.266,0.266,0.266}
\begin{tikzpicture}[line cap=round,line join=round,>=triangle 45,x=1cm,y=1cm]
\draw [line width=2pt] (0,0) circle (3cm);
\draw [rotate around={0:(-0.01,0)},line width=2pt,color=zzttff,fill=gray!20] (-0.01,0) ellipse (2.79576075716924cm and 1.2313318851258266cm);
\draw [line width=2pt] (0.80,1.18)-- (-0.80,-1.18);
\begin{scriptsize}
\draw[color=xdxdff] (-1.74,2.76) node {$i$};
\draw [fill=xdxdff] (-1.73,-2.45) circle (2pt);
\draw[color=xdxdff] (-1.73,-2.76) node {$j$};
\draw [fill=xdxdff] (-2.60,-1.49) circle (2pt);
\draw[color=xdxdff] (-2.74,-1.96) node {$h$};
\draw [fill=magenta] (-1.6,0) circle (2pt);
\draw [fill=magenta] (0.71,0.31) circle (2pt);
\draw[line width=2pt,color=pink] (0.71,0.31) -- (-0.28,-1.23);
\draw [fill=ccqqqq] (1,0) circle (2pt);
\draw [fill=xdxdff] (1.8,-0.94) circle (2pt);
\draw[color=xdxdff] (2.2,-1.2) node {$Y$};
\draw [fill=xdxdff] (-2.28,-0.71) circle (2pt);
\draw [fill=xdxdff] (2.07,-2.18) circle (2pt);
\draw[color=xdxdff] (2.38,-2.46) node {$l$};
\draw [fill=xdxdff] (2.21,2.024) circle (2pt);
\draw[color=xdxdff] (2.38,2.46) node {$k$};
\draw[line width=2pt,color=pink] (-2.28,-0.71) -- (-1.6,0) -- (0,0);
\draw[line width=2pt,color=pink] (2.42,0.61) -- (1.45,-0.54);
\draw[line width=2pt,color=pink] (-1.6,0) -- (-1.77,0.96);
\draw[line width=2pt,color=pink] (1,0) -- (0.43,0.63);
\draw[line width=2pt,color=black] (1,0) -- (1.8,-0.94);
\draw[thick,blue,dashed] plot [smooth] coordinates {(-1.90,2.33) (-0.8,0) (-1.73,-2.45)};
\node at (-0.8,0){\(\bullet\)};
\node at (-0.5,0.2){\small\(\phi_{ij}\)};
\node at (-2.2,-0.3){\small\(\psi_{h}^{MN}\)};
\draw[thick,bend right,blue,dashed] (-2.60,-1.49) edge (-2.28,-0.71);
\node[color=xdxdff] at (-2,-1){\(M\)};
\node[color=xdxdff] at (-2.6,-0.8){\(N\)};
\draw[thick,blue,dashed] plot [smooth] coordinates {(2.07,-2.18) (1,0)
(2.21,2.024)};
\node at (1,0){\(\bullet\)};
\node at (1.5,0.2){\small\(\phi_{ij}^{Y}\)};
\draw [fill=ududff] (-1.77,0.96) circle (2pt);
\draw [fill=ududff] (2.42,0.61) circle (2pt);
\draw [fill=xdxdff] (-0.80,-1.18) circle (2pt);
\draw [fill=xdxdff] (-0.28,-1.23) circle (2pt);
\draw[color=xdxdff] (-1,-1.6) node {$X$};
\draw [fill=xdxdff] (-1.90,2.33) circle (2pt);
\draw[color=xdxdff] (-1.74,2.76) node {$i$};
\draw [fill=xdxdff] (-1.73,-2.45) circle (2pt);
\end{scriptsize}
\end{tikzpicture}
\caption{Illustration of the three dimensional bordism of a disk with two defects(black solid lines) \(X\) and \(Y\). The dashed blue lines represent objects in \(\mathcal{C}\), whereas each pink solid line is labeled by a Frobenius algebra.}
\label{three bordism}
\end{figure}

\noindent The TFT construction of \(M_\Sigma\) is obtained by several steps, and is illustrated in figure \ref{three bordism}. 
\noindent First, choose a dual triangulation on \(\Sigma\) and label each edge of the triangulation with a Frobenius algebra in \(\Obj(\mathcal{C})\). Then, fill the ribbons in \(M_\Sigma\) starting from the boundary, and label them with objects in $\mathcal{C}$. The intersection points of ribbons on the interior of $\iota(\Sigma)$ correspond to the bulk fields while those on the boundary of $\iota(\Sigma)$ correspond to the boundary fields. others CFT quantities are list in table \ref{dictionary between CFT quantity and algebraic notion}.  The correlators associated to such a ribbon graph is independent of the triangulation\cite{Fuchs:2002cm}. Since we have doubled the worldsheet, there will be two labels before and after the piercing which roughly correspond to the left and right movers of the CFT. A simple interpretation of \(\Sigma\) is a surface operator in \(M_X\)\cite{Kapustin:2010if}.

\begin{table}[ht]
\centering
\begin{tabular}{|c|c|}\hline
Physical quantities&Algebraic notions \\\hline
Boundary conditions \(M\)&\(A\)-modules  \\\hline
Defect lines \(X\)&\(A\)-bimodules  \\\hline
Boundary fields \(\psi_{h}^{M,N}\)& \(\Hom(U_h\otimes M,N)\) \\\hline
Bulk fields \(\phi_{ij}\) &\(\Hom(U_i\otimes^+ A\otimes^-U_j,A)\)\\\hline
Disorder fields \(\phi_{kl}^Y\) &\(\Hom(U_k\otimes^+ Y\otimes^-U_l,A)\)\\\hline
\end{tabular}
\caption{Dictionary between CFT quantities and algebraic structures.}
\label{dictionary between CFT quantity and algebraic notion}
\end{table}
The triangulation operation can be understood as gauging a symmetry. \cite{Carqueville:2012dk, Brunner:2013xna, Bhardwaj:2017xup, Brunner:2014lua,Komargodski:2020mxz}.\footnote{We thank Jin Chen for pointing this out to us.} Consider an diagonal RCFT \(\mathcal{T}\) living on \(\Sigma\). We have proved in section \ref{diagonal and permutation model} that the fusion category of the TDLs are given by \(\mathcal{C}\) by forgetting the braiding structure. Since TDLs are a generalized notion of symmetry, we can gauge a non-anomalous subpart \(A\). This is achieved by inserting a fine-enough mesh of an algebra object \(A\) in \(\mathcal{C}\). By "Fine-enough mesh", we mean that it can be obtained as the dual graph of a triangulation of \(\Sigma\). Drawing from the analogy of ordinary group symmetry cases,  \(A\) is argued to be the  symmetric special Frobenius algebra\cite{Bhardwaj:2017xup}. Consequently, the generalized symmetries/TDLs of the gauged theory is represented by bimodules \(\mathcal{C}_{A|A}\). We will work out the bimodules of our main example in section \ref{tftpotts}.

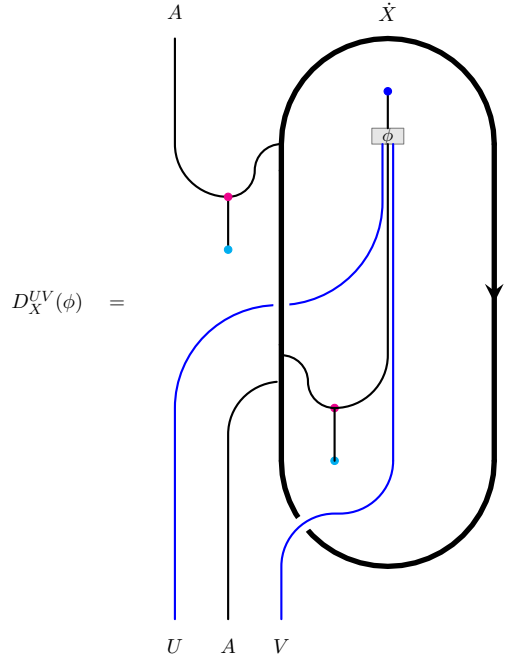
\begin{figure}[h!t]
\centering
\definecolor{ududff}{rgb}{0.30196078431372547,0.30196078431372547,1}
\begin{tikzpicture}[line cap=round,line join=round,>=triangle 45,x=1cm,y=1cm,scale=0.7, every node/.style={scale=0.7}]
\node at (-6,0){\(D_{X}^{UV}(\phi)\quad=\)};
\node at (0,5.5){\(\dot{X}\)};
\draw [shift={(0,3)},line width=2pt,cross]  plot[domain=0:3.141592653589793,variable=\t]({1*2*cos(\t r)+0*2*sin(\t r)},{0*2*cos(\t r)+1*2*sin(\t r)});
\draw [shift={(0,-3)},line width=2pt]  plot[domain=3.141592653589793:6.283185307179586,variable=\t]({1*2*cos(\t r)+0*2*sin(\t r)},{0*2*cos(\t r)+1*2*sin(\t r)});
\draw [line width=2pt,
        decoration={markings, mark=at position 0.5 with {\arrow{stealth}}},
        postaction={decorate}] (2,3)-- (2,-3);
\draw[thick] (-3,1) -- (-3,2) arc (-90:-180:1) -- (-4,5);
\draw[thick] (-3,2) arc (-90:0:0.5) arc (-180:-270:0.5);
\filldraw[magenta] (-3,2) circle (2pt);
\filldraw[cyan] (-3,1) circle (2pt);
\draw[gray,fill=gray!20] (-0.3,3) rectangle ++(0.6,0.3);
\node at (0,3.15){\small\(\phi\)};
\filldraw[magenta] (-1,-2) circle (2pt);
\filldraw[cyan] (-1,-3) circle (2pt);
\draw[thick] (-1,-3) -- (-1,-2) arc (-90:-180:0.5) arc (0:90:0.5);
\draw[thick] (-1,-2) arc (-90:0:1) -- (0,3);
\draw[thick,cross,blue] (-2,-6) -- (-2,-5) arc (-180:-270:1)-- (-0.9,-4) arc (-90:0:1) ;
\draw[thick,blue] (0.1,-3) -- (0.1,3);
\draw[thick] (-3,-6) -- (-3,-2.5) arc (-180:-265:1);
\draw[thick,blue] (-4,-6) -- (-4,-2) arc (-180:-270:1.95) arc (-90:0:1.95) -- (-0.1,3);
\draw [line width=2pt,cross] (-2,2.5)-- (-2,-0.8);
\draw [line width=2pt] (-2,2.5) -- (-2,3);
\draw [line width=2pt] (-2,-0.8) -- (-2,-3);
\draw[thick] (0,3.3) -- (0,4);
\filldraw[blue] (0,4) circle (2pt);
\node at (-4,5.5){\(A\)};
\node at (-4,-6.5){\(U\)};
\node at (-3,-6.5){\(A\)};
\node at (-2,-6.5){\(V\)};
\end{tikzpicture}
\caption{Defects acting on bulk fields} 
\label{defect acting on bulk field}
\end{figure}

After making a connection between the CFT and TFT quantities, we have the precise description to consider a bulk field insertion encircled by a defect. The action can map a bulk field to another with the same conformal weight, but the internal or extended charges can be different, because we only require the TDLs preserve the stress tensor. This actually define a map between the defects and the endomorphisms of bulk fields. For a given bulk field \(\phi\in \Hom(U_i\otimes^+ A\otimes^-U_j,A)\) and a defect \(X\in \mathcal{C}_{A|A}\), we can define the endomorphism in terms of \(X\) by zooming in the bulk fields and the defects in figure \ref{three bordism}, and we get graph \ref{defect acting on bulk field}. 

\noindent The graph can be represented as,
\begin{equation}
\begin{aligned} \phi\mapsto D_X^{U V}(\phi):= & \left(\identity_A \otimes \tilde{e}_{\dot{X}}\right) \circ\left[\left(\left[\identity_A \otimes \rho_{\ell} \otimes(\varepsilon \circ \phi)\right] \circ\left[(\Delta \circ \eta) \otimes c_{U, \dot{X}} \otimes \identity_A \otimes \identity_V\right]\right) \otimes \identity_{\dot{X} ^{*}}\right] \\ & \circ\left[\identity_U \otimes\left(\left[\left(\rho_{r} \circ\left(\rho_{\ell} \otimes \identity_A\right)\right) \otimes \identity_A\right] \circ\left[\identity_A \otimes \identity_{\dot{X}} \otimes(\Delta \circ \eta)\right]\right) \otimes \identity_V \otimes \identity_{\dot{X} ^{*}}\right] \\ & \circ\left[\identity_U \otimes \identity_A \otimes\left(\left[c_{\dot{X}, V}^{-1} \otimes \identity_{\dot{X} ^{*}}\right] \circ\left[\identity_V \otimes i_{\dot{X}}\right]\right)\right].
\end{aligned}
\end{equation}

\begin{figure}[ht]
\centering
\begin{tikzpicture}[line cap=round,line join=round,>=triangle 45,x=1cm,y=1cm,scale=0.5, every node/.style={scale=0.5}]
\node at (-4.8,-0.2){\Large{\(D^{U_iU_i}_{X_m}(\phi)\)}};
\node at (-3.5,-0.2){\(=\)};
\draw[blue,line width=1pt](0,-3) -- (0,0);
\draw [cross,rotate around={0:(-0.01,0)},line width=1pt,
        decoration={markings, mark=at position 0.8 with {\arrow{stealth}}},
        postaction={decorate}] (-0.01,0) ellipse (2.79576075716924cm and 1.2313318851258266cm);
\draw[blue,line width=1pt,cross,
        decoration={markings, mark=at position 0.5 with {\arrow{stealth}}},
        postaction={decorate}](0,0) -- (0,2.5);
\draw[blue,line width=1pt](0,-1)--(0,1);
\node at (0.5,-0.2){\(\phi\)};
\node at (2.5,-1){\(X_m\)};
\node at (3.5,-0.2){\(=\)};
\node at (4.5,-0.2){\Large{\(\frac{S_{mi}}{S_{i0}}\)}};
\draw[blue,line width=1pt,
        decoration={markings, mark=at position 0.8 with {\arrow{stealth}}},
        postaction={decorate}](5.5,-3) -- (5.5,2.5);
\node at (6,-0.2){\(\phi\)};
\end{tikzpicture}
\caption{Defects acting on bulk fields in diagonal case.}
\label{diagonal acting on bulk fields}
\end{figure}
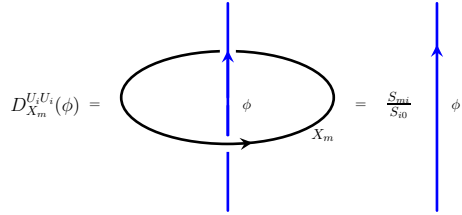

\noindent When reducing to diagonal case, namely \(A=1\), \(A\) lines become invisible, and the graph will be simplified significantly, as depicted in figure \ref{diagonal acting on bulk fields}. The evaluation of the graph is precisely given by equation \eqref{eq:Verlindeline} \cite{Bakalov2000LecturesOT}.

Different defects loop will give different endomorphisms, i.e., different representation matrices in the space of conformal primaries\cite{Fuchs:2007vk}. There is actually an isomorphism between the complexified Grothendieck ring of the category of defects and the endormorhphism of bulk fields, namely the bijection,
\begin{align}\label{iso endo of bulk fiels and defects}
 K_{0}(\mathcal{C}_{A|A}) \otimes_{\Z} \CC \eqqcolon \mathcal{F} \xrightarrow{\sim}\mathcal{E}\coloneqq \bigoplus\limits_{i,j\in \mathcal{I}} \End_{\CC}\left(\Hom_{A|A}\left(U_i\otimes^+ A\otimes^-U_j,A\right)\right).  
\end{align}
preserves the following algebraic structure,
\begin{equation}
    D_{A} =\identity; \quad D_{X}\circ D_{Y} = D_{X\otimes_A Y}.
\end{equation}
Moreover, we have 
\begin{equation}\label{number of bimodules}
    \text{dim}_{\CC}(\mathcal{F}) = \text{dim}_{\CC}(\mathcal{E}) \equiv \sum_{ij\in \mathcal{I}}(z_{ij})^2,
\end{equation}
where \(z_{ij}=\text{dim}\Hom(U_i\otimes^+ A\otimes^-U_j,A)\). This formula tells us the number of simple defects if we know all the bulk fields. In the three-state Potts model, we get \(16\) simple TDLs.

We choose a basis  \(\ket{\phi_{\alpha}^{(ij)}}\) of \(\Hom(U_i\otimes^+ A\otimes^-U_j,A)\) and \(\bra{\phi_{\beta}^{(ij)}}\) as it's dual with \(\alpha,\, \beta = 1,\ldots,z_{ij}\). Then the defect \(X_{k}\) can be represented as a matrix because of the isomorphism \eqref{iso endo of bulk fiels and defects},
\begin{align}
D_{X_{k}} \ket{\phi_{\alpha}^{(ij)}} = d_{k}^{ij;\alpha,\beta} \ket{\phi_{\beta}^{(ij)}}.
\label{matrixRep}
\end{align}
We can reformulate the defect Cardy condition. Consider the twined partition function, 
\begin{equation}
\begin{aligned}
Z^{k} &=\sum_{ij\in\mathcal{I}}\sum_{\alpha,\beta}^{z_{ij}} \sum_{\text{descendants}}\left\langle \phi_{\alpha}^{(ij)}\right| q^{L_0-\frac{c}{24}}\bar{q}^{\bar{L}_0-\frac{\bar{c}}{24}}D_{X_{k}} \left| \phi_{\alpha}^{(ij)}\right\rangle\\
&=\sum_{ij\in\mathcal{I}}\sum_{\alpha,\beta}^{z_{ij}}d_{k}^{ij;\alpha,\beta}\chi_{i\alpha}(\tau)\bar{\chi}_{j\beta}(\bar{\tau})\delta_{\alpha,\beta}\\
&=\sum_{ij\in\mathcal{I}}\sum_{\alpha}^{z_{ij}} d_{k}^{ij;\alpha,\alpha}\chi_{i\alpha}(\tau)\bar{\chi}_{j\alpha}(\bar{\tau}),
\end{aligned}
\end{equation}
where \(\chi_{i\alpha}(\tau)=\chi_{i\beta}(\bar{\tau})\) as Virasoro characters. Under the \(S\) transformation, we get the defect partition function,
\begin{equation}
\begin{aligned}
Z_{k} &= \sum_{jl}\sum_{ij\in\mathcal{I}}\sum_{\alpha}^{z_{ij}} d_{k}^{ij;\alpha,\alpha}\chi_{i\alpha}(\tau)\bar{\chi}_{j\alpha}(\bar{\tau})S_{ik}S^*_{jl}\\
&=\sum_{kl}M_{kl}\chi_{k}(\tau)\chi_{l}(\bar{\tau}),\; M_{kl}\in \Z_{\geq 0}
\end{aligned}
\end{equation}
The defect Cardy condition is then,
\begin{align}
\sum_{i,j}\Tr d_{k}^{ij;\alpha,\beta}S_{ik}S_{jl} = M_{kl},\quad  M_{kl}\in \Z_{\geq 0},
\label{defectCardyII}
\end{align}
where the \(\tr\) operator is the trace over the matrix with entries \(\alpha\) and \(\beta\). A parallel derivation with simpler notation is near equation \eqref{defectCardyIII}. The intricate index here is helpful for clarifying the physical meaning within the TFT construction.

\subsection{three-state Potts model again}\label{tftpotts}
In order to verify the TDLs and fusion rules we have found in section \ref{block diagonal}, we would like to compute the TDLs and their fusion rule again by the simple current construction, where the corresponding algebra is constructed from invertible objects\cite{Fuchs:2004dz}. 

The MTC for three-state Potts model is \(\mathfrak{su}(2)_{4} \times \text{Lee-Yang}\), whose simple objects are:
\begin{align}
\mathcal{I} = \{1,u,f,v,w,\hat{1},\hat{u},\hat{f},\hat{v},\hat{w}\}.
\end{align}
Their fusion rules can be derived from table \ref{fusion su(2)_4} and \eqref{fusion of c65}.
\begin{align}
\hat{1}\otimes x = \hat{x}, \;\; x\in \mathfrak{su}(2)_4;\quad \hat{1}\otimes \hat{1} = 1\oplus \hat{1}.
\label{fusion of c65}
\end{align}

\begin{table}[h!t]
\centering
\begin{tabular}{|c|c|c|c|c|c|}\hline
$\otimes$ & $1$ &$u$&$f$&$v$&$w$\\\hline
$1$& $1$&$u$&$f$&$v$&$w$\\\hline
$u$&$u$ &$1\oplus f$&$u\oplus v$&$f\oplus w$&$v$\\\hline
$f$ & $f$ & $u\oplus v$ & $1\oplus f\oplus w$&$u\oplus v$&$f$\\\hline
$v$& $v$&$f\oplus w$&$u+v$&$1\oplus f$&$u$\\\hline
$w$&$ w$&$v$&$f$&$u$&$1$\\\hline
\end{tabular}
\caption{The fusion algebra of \(\mathfrak{su}(2)_{4}.\)}
\label{fusion su(2)_4}
\end{table}

\begin{figure}[ht]
\centering
\begin{tikzpicture}[scale=0.5, every node/.style={scale=0.5}]
\draw[-,thick] (0,0) -- (0,4);
\draw[-,thick] (-2,0) -- (-2,1) arc (-180:-270:2);
\node at (-2,-0.5){\(B\)};
\node at (0,-0.5){\(B\otimes U\)};
\node at (0,4.5){\(B\otimes U\)};
\node at (1,2){\(=\)};
\draw[-,thick] (2,0) -- (2,2.5) arc (-180:-360:0.5) -- (3,0);
\draw[-,thick] (2.5,3) -- (2.5,4);
\draw[-,thick] (4,0) -- (4,4);
\node at (2,-0.5){\(B\)};
\node at (3,-0.5){\(B\)};
\node at (4,-0.5){\(U\)};
\node at (4,4.5){\(U\)};
\node at (2.5,4.5){\(B\)};
\filldraw[magenta] (2.5,3) circle (2pt);
\end{tikzpicture}
\caption{Induced module structure.}
\label{induced module structure}
\end{figure}
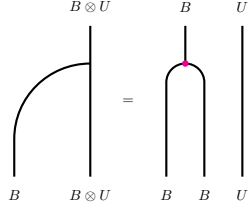

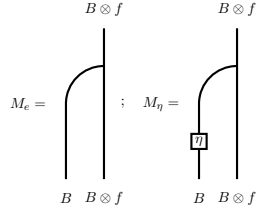
\begin{figure}[ht]
\centering
\begin{tikzpicture}[scale=0.5, every node/.style={scale=0.5}]
\node at (-2,2){\(M_e =\)};
\draw[-,thick] (0,0) -- (0,4);
\draw[-,thick] (-1,0) -- (-1,2) arc (-180:-270:1);
\node at (-1,-0.5){\(B\)};
\node at (0,-0.5){\(B\otimes f\)};
\node at (0,4.5){\(B\otimes f\)};
\node at (0.5,2){;};
\node at (1.5,2){\(M_{\eta} =\)};
\draw[-,thick] (2.5,0) -- (2.5,0.8);
\draw[-,thick] (2.3,0.8) rectangle ++(0.4,0.4);
\node at (2.5,1){\(\eta\)};
\draw[-,thick] (2.5,1.2) -- (2.5,2) arc (-180:-270:1);
\draw[-,thick] (3.5,0) -- (3.5,4);
\node at (2.5,-0.5){\(B\)};
\node at (3.5,-0.5){\(B\otimes f\)};
\node at (3.5,4.5){\(B\otimes f\)};
\end{tikzpicture}
\caption{Decomposition of \(M_f\)}
\label{Decomposition of M_f}
\end{figure}

\noindent From the table, we can see that \(1\) and \(w\) are invertible objects. So the Frobenius algebra of this model is \(B = 1 \oplus w\). The first step to find \(B\)-bimodules is to construct left \(B\)-modules, which can be constructed by the \(B\)-induced modules\cite{Fuchs:2004dz}. The induced modules are given by
\begin{align}
M_{U} \coloneqq \operatorname{Ind}_B(U) = (B\otimes U, m\otimes \identity_U)
\end{align}
where $U\in \Obj(C)$ and $m$ is the algebra product. Pictorially, the module structure is given by figure \ref{induced module structure}.
These induced modules are not necessarily simple in general. To find the simple ones, the following reciprocity property will help us to simplify the computation. Let \(U \in \Obj(C)\) and \(M\) be a left \(B\)-module. There is a natural isomorphism between the MTC and the category of left \(B\)-modules,
\begin{align}
    \Hom(U,\dot{M}) &\xrightarrow{\cong}\Hom_{B}(\text{Ind}_{B}(U),M)\\ 
    \varphi &\mapsto \rho_{M}\circ (\identity_B\otimes \varphi).
\end{align}
It follows that \(\text{dim}_{\CC} \Hom_B (\text{Ind}(U),\text{Ind}(U)) = \text{dim}_{\CC} \Hom(B\otimes U ,U)\). We find that \(M_{U}\), \(U=1,u,\hat{1},\hat{u}\) are simple, whereas \(M_{f}\) and \(M_{\hat{f}}\) are not, see table \ref{Dimension of the induced modules}. 
\begin{table}[h!t]
\centering
\begin{tabular}{|c|c|c|c|c|c|c|c|c|c|c|}\hline
$U\in\Obj(\mathcal{C})$ & $1$ & $u$ & $f$ & $v$ & $w$ & $1$ & $\hat{u}$ & $\hat{f}$ & $\hat{v}$ & $\hat{w}$\\\hline
$\text{dim}_{\CC}(M_{U})$ &$1$ &$1$  &$2$  &$0$  &$0$  &$1$ &$1$  &$2$  &$0$  &$0$\\\hline
\end{tabular}
\caption{Dimension of the induced modules}
\label{Dimension of the induced modules}
\end{table}
\noindent A convenient but not canonical decomposition of \(M_{f}\) is $M_f=M_e\oplus M_\eta$, where \(M_{\eta}\) is given by composing \(\eta\otimes \identity\) before \(B\) acting on \(B\otimes f\), see figure \ref{Decomposition of M_f}. Similarly, we have \(M_{\hat{f}}=\text{Ind}_{B}(\hat{f}) \cong \hat{M}_{\hat{e}}\oplus \hat{M}_{\hat{\eta}}\).

\begin{figure}[ht]
\centering
\begin{tikzpicture}[scale=0.5, every node/.style={scale=0.5}]
\node at (0,-0.5){\(\dot{M}_U\)};
\node at (0,5.5){\(\dot{M}_U\)};
\draw[-,thick] (1,0) -- (1,1) arc (0:90:1) arc (-90:-180:1);
\draw[-,thick,cross] (0,0) -- (0,5);
\draw[-,thick] (-1.2,3) rectangle ++(0.4,0.4);
\node at (-1,3.2){\(\psi\)};
\draw[-,thick] (-1,3.4) arc (-180:-270:1);
\draw[-,thick] (3,0)--(3,3) arc (-180:-270:0.5) -- (3.5,5);
\node at (2,2){\(=\)};
\draw[-,thick] (4.8,1.5) rectangle ++(0.4,0.4);
\node at (5,1.7){\(\psi\)};
\draw[-,thick] (5,0)--(5,1.5);
\draw[-,thick] (5,1.9)--(5,2) arc (0:90:1.5);
\draw[-,thick,cross] (4,0) -- (4,5);
\node at (1,-0.5){\(B\)};
\node at (3,-0.5){\(B\)};
\node at (4,-0.5){\(U\)};
\node at (5,-0.5){\(B\)};
\node at (3.5,5.5){\(B\)};
\node at (4,5.5){\(U\)};
\end{tikzpicture}
\caption{Right module strutures in \ref{bimodule pair}}
\label{Right module struture of the pair}
\end{figure}
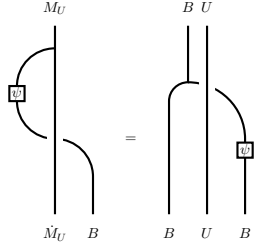

\noindent  Now we can list all the \(8\) simple left $B$-modules of three-state Potts model.
\begin{align}\label{left module index set}
\mathcal{J}_{B} = \{1,u,e,\eta,\hat{1},\hat{u},\hat{e},\hat{\eta}\}
\end{align}
For an induced left module, say \(\text{Ind}_{B}(U)\), there are two natural ways to endow it right module structures with the help of braiding and inverse braiding respectively,
\begin{equation}
    \rho_r^{+} \coloneqq (m\otimes \identity_U )\circ (\identity_B\otimes c_{U,B}),\quad \rho_r^{-} \coloneqq (m\otimes \identity_U )\circ (\identity_B \otimes c^{-}_{U,B}).
\end{equation}
In fact, the above discussion defines a tensor functor \(\alpha^{\pm}\) from the MTC \(\mathcal{C}\) to the bimodule category \(\mathcal{C}_{B|B}\). This is known as alpha induction functor in the study of sub-factor theory which is originally used to construct modular invariants\cite{Bockenhauer:1998ca,Bockenhauer:1998in,Bockenhauer:1998ef,Bockenhauer:1999wt}. We pick the first one, and the right module structure is then given by figure \ref{Right module struture of the pair}. 
Note that \(B\) is commutative, and then the isomorphism classes of simple \(B\)-bimodules can be labeled by pairs consisting of an element from \(\mathcal{J}_B\) and an algebra automorphism\cite{Frohlich:2006ch}.
\begin{align}
\mathcal{K}_{B|B} = \{(k,\psi)|k \in \mathcal{J}_B, \psi\in \Aut(B)\}.
\label{bimodule pair}
\end{align}
Here, \(\Aut(B) = \{e,\eta\} \cong \Z_2\) with 
\begin{align}
e=\identity_{B};\quad \eta=\identity_1 \oplus (-\identity_{w})
\end{align}
 The number of simple bimodule is \(|\mathcal{J}_{B}| \times |\text{Aut}(B)|\), and this is consistent with the calculation by imposing equation \eqref{number of bimodules}.

\noindent We end this section by calculating the fusion rules of these bimodules and comparing with the ones in section \ref{block diagonal}. Since \(\alpha^{+}\) is a tensor functor, we have 
\begin{equation}
\left[\alpha_B^{+}\left(U\right)_\phi\right] \otimes_{B}\left[\alpha_B^{+}\left(V\right)_\psi\right]=\left[\alpha_B^{+}\left(U \otimes V\right)_{\phi \psi}\right].
\end{equation}
Using this formula, we can write down the following fusion rules:
\begin{equation}
\begin{aligned}
& (1, \phi) (1, \psi)=(1, \phi \psi), \quad(1, \phi)(u, \psi)=(u, \phi \psi), \\
& (u, \phi) (u, \psi)=(1, \phi \psi)\oplus(e, \phi \psi)\oplus(w, \phi \psi), \\
& ((e, \phi)\oplus(\eta, \phi)) (u, \psi)=2(u, \phi \psi) .
\end{aligned}
\label{fusion derived from tenor functor property}
\end{equation}
where we omitted the fusion product $\otimes_B$.
It follows that 
\begin{equation}
    (1, \phi)(u, \psi) = (e, \phi)(u, \psi)=(\eta, \phi)(u, \psi)=(u, \phi \psi)
\end{equation}
\begin{table}[ht]
\centering
\begin{tabular}{|c|c|c|c|c|c|c|c|c|}\hline
$(k,\psi)$ & ${(1,e)}$ & $(e,e)$ & $(\eta,e)$ &$(u,e)$ & $(1,\eta)$ & $(e,\eta)$ & $(\eta,\eta)$ & $(u,\eta)$\\\hline
Simple TDL &$1$ &$\eta$  &$\bar{\eta}=\eta^2$  &$N$  &$C\eta$  &$C$ &$C\eta^2$ &$N^\prime=CN$ \\\hline
\(S_3\) element&$\identity$ &$(123)$  &$(132)$  &  &$(12)$  &$(13)$ &$(23)$  &\\\hline
\end{tabular}
\caption{We denote these fusion category as \(\alpha_B^{+}(\mathfrak{su}(2)_4)\). The notation in the second line is the one used in section \ref{block diagonal}, with fusion rules \eqref{fusionTP}. }
\label{List of the defects}
\end{table}

\noindent These bimodules can be identified with simple TDLs in section \ref{block diagonal} as shown in Table \ref{List of the defects}. Moreover, we can see six of them can be identified with elements of the permutation group $S_3$. The fusion rules of \(\hat{x}\) with \(x\in \{1,u,e,\eta\}\), can be computed similarly. We have
\begin{equation}
    (x,\phi)\otimes_{B}(\hat{1},e) = (\hat{1},e)\otimes_{B}(x,\phi) = (\hat{x},\phi)\quad \forall x\in \{1,u,e,\eta\} \text{ and } \phi\in \text{Aut}(B).
    \label{fusionwithhat}
\end{equation}
$(\hat{1},e)$ corresponds to the $W$ line.
Using \eqref{fusionwithhat}, we can get another \(8\) defects \(WX=XW\) for all \(X\in \alpha_B^{+}\left(\mathfrak{su}(2)_4\right)\) and the corresponding fusion rules. We denote the whole category as  \(\alpha_B^{+}\left(\mathfrak{su}(2)_4\right)\otimes_{B} \alpha_B^{+}\text{(Lee-Yang)}\).
\section{Conclusions}\label{conclusion}
We proved that in diagonal and permutation diagonal minimal model, Verlinde lines are complete set of the simple TDLs. For block-diagonal minimal models, we proposed a recipe for finding all the simple TDLs using fusion rules consistency and tested it on the 3-state Potts model. Inferring the fusion rules will soon become intricate for minimal models $\mathcal{M}(p,p+1)$ with higher $p$ since the number of primaries increase fast.

A more interesting model for running this procedure to find TDLs is $\mathcal{M}(12,5)$, which is the first (lowest central charge) $E$-type minimal model. This is because explicit connection between $E$-series and the corresponding $A$-series model have not been established via the anyon condensation. Yet more interesting applications are in fermionic theories, especially those with supersymmetry, where the mathematical understanding of TDLs is still underdeveloped.    We are also hoping to apply this procedure to higher dimensions, say to $3$-dimensional CFT. The main difficulty is, our method relies heavily on the modular properties which have no apparent generalizations in higher dimensional theories. A novel mathematical structure of higher categories may serve as a tool.

\section*{Acknowledgement}
The authors would like to thank Jin Chen, Ling-Yan Hung, Youran Sun for valuable discussions and comments. Special thanks go to Chi-Ming Chang, whose initial guidance and advice are greatly appreciated.

\appendix

\section{The coefficient matrix \texorpdfstring{$M$}{M} in the 3 state Potts model}
A twisted partition function with a defect $\mathcal{L}$ in the three state Potts model can be written as
\begin{equation}
    Z_{\mathcal{L}}=M_{ij}\chi_i\chi_j.
\end{equation}
There are $10$ possible characters $\chi_i$ appearing in the right hand side. We can label the characters in the following order: $(\chi_1,\chi_Y,\chi_\epsilon,\chi_X,\chi_{Z(Z^*)},\chi_{\sigma(\sigma^*)},\chi_{2,4},\chi_{4,4},\chi_{2,2},\chi_{2,3})$. Here the first six characters are labeled by the corresponding primary fields in 3 state Potts model, while the latter 4 characters are labeled by Kac indices in the corresponding $c=\frac45$ diagonal minimal model. The modular transformation necessarily brings out characters in the diagonal models.  We use $M_{ij}$ to denote the coefficient of the product of the $i$-th holomorphic character and the $j$-th antiholomorphic one. For example, $M_{23}$ corresponds to $\chi_Y\bar{\chi}_\epsilon$. 

$M_{ij}$'s are not all independent, having many relations from the modular transformation. One of the relation is
\begin{equation}
    \begin{aligned}
 M_{11}=M_{12}=M_{21}=M_{22}.
    \end{aligned}
\end{equation}
Solving these relations leaves only $10$ independent $M_{ij}$'s. A general matrix $M$ is a linear combination of the following $10$ matrices:
\begin{equation}
\begin{aligned}
    &M^1=\begin{pmatrix}
        1&1&0&0&0&0&0&0&0&0\\
        1&1&0&0&0&0&0&0&0&0\\
        0&0&1&1&0&0&0&0&0&0\\
        0&0&1&1&0&0&0&0&0&0\\
        0&0&0&0&2&0&0&0&0&0\\
        0&0&0&0&0&2&0&0&0&0\\
        0&0&0&0&0&0&0&0&0&0\\
        0&0&0&0&0&0&0&0&0&0\\
        0&0&0&0&0&0&0&0&0&0\\
        0&0&0&0&0&0&0&0&0&0\\

    \end{pmatrix},\;
    M^2=\begin{pmatrix}
        0&0&1&1&0&0&0&0&0&0\\
        0&0&1&1&0&0&0&0&0&0\\
        1&1&1&1&0&0&0&0&0&0\\
        1&1&1&1&0&0&0&0&0&0\\
        0&0&0&0&0&2&0&0&0&0\\
        0&0&0&0&2&2&0&0&0&0\\
        0&0&0&0&0&0&0&0&0&0\\
        0&0&0&0&0&0&0&0&0&0\\
        0&0&0&0&0&0&0&0&0&0\\
        0&0&0&0&0&0&0&0&0&0\\
        
    \end{pmatrix},\\
    &M^3=\begin{pmatrix}
        0&0&0&0&1&0&0&0&0&0\\
        0&0&0&0&1&0&0&0&0&0\\
        0&0&0&0&0&1&0&0&0&0\\
        0&0&0&0&0&1&0&0&0&0\\
        1&1&0&0&1&0&0&0&0&0\\
        0&0&1&1&0&1&0&0&0&0\\
        0&0&0&0&0&0&0&0&0&0\\
        0&0&0&0&0&0&0&0&0&0\\
        0&0&0&0&0&0&0&0&0&0\\
        0&0&0&0&0&0&0&0&0&0\\

    \end{pmatrix},\;
    M^4=\begin{pmatrix}
        0&0&0&0&0&1&0&0&0&0\\
        0&0&0&0&0&1&0&0&0&0\\
        0&0&0&0&1&1&0&0&0&0\\
        0&0&0&0&1&1&0&0&0&0\\
        0&0&1&1&0&1&0&0&0&0\\
        1&1&1&1&1&1&0&0&0&0\\
        0&0&0&0&0&0&0&0&0&0\\
        0&0&0&0&0&0&0&0&0&0\\
        0&0&0&0&0&0&0&0&0&0\\
        0&0&0&0&0&0&0&0&0&0\\

    \end{pmatrix},\\
    &M^5=\begin{pmatrix}
        0&0&0&0&0&0&1&1&0&0\\
        0&0&0&0&0&0&1&1&0&0\\
        0&0&0&0&0&0&0&0&1&1\\
        0&0&1&1&0&0&0&0&1&1\\
        0&0&0&0&0&0&2&2&0&0\\
        0&0&0&0&0&0&0&0&2&2\\
        0&0&0&0&0&0&0&0&0&0\\
        0&0&0&0&0&0&0&0&0&0\\
        0&0&0&0&0&0&0&0&0&0\\
        0&0&0&0&0&0&0&0&0&0\\

    \end{pmatrix},\;
    M^6=\begin{pmatrix}
        0&0&0&0&0&0&0&0&1&1\\
        0&0&0&0&0&0&0&0&1&1\\
        0&0&0&0&0&0&1&1&1&1\\
        0&0&1&1&0&0&1&1&1&1\\
        0&0&0&0&0&0&0&0&2&2\\
        0&0&0&0&0&0&2&2&2&2\\
        0&0&0&0&0&0&0&0&0&0\\
        0&0&0&0&0&0&0&0&0&0\\
        0&0&0&0&0&0&0&0&0&0\\
        0&0&0&0&0&0&0&0&0&0\\

    \end{pmatrix},\\
    &M^7=\begin{pmatrix}
        0&0&0&0&0&0&1&1&0&0\\
        0&0&0&0&0&0&1&1&0&0\\
        0&0&0&0&0&0&0&0&1&1\\
        0&0&1&1&0&0&0&0&1&1\\
        0&0&0&0&0&0&2&2&0&0\\
        0&0&0&0&0&0&0&0&2&2\\
        0&0&0&0&0&0&0&0&0&0\\
        0&0&0&0&0&0&0&0&0&0\\
        0&0&0&0&0&0&0&0&0&0\\
        0&0&0&0&0&0&0&0&0&0\\
    \end{pmatrix}^T,\;
    M^8=\begin{pmatrix}
        0&0&0&0&0&0&0&0&1&1\\
        0&0&0&0&0&0&0&0&1&1\\
        0&0&0&0&0&0&1&1&1&1\\
        0&0&1&1&0&0&1&1&1&1\\
        0&0&0&0&0&0&0&0&2&2\\
        0&0&0&0&0&0&2&2&2&2\\
        0&0&0&0&0&0&0&0&0&0\\
        0&0&0&0&0&0&0&0&0&0\\
        0&0&0&0&0&0&0&0&0&0\\
        0&0&0&0&0&0&0&0&0&0\\
    \end{pmatrix}^T,
\end{aligned}
\end{equation}
\begin{equation}
    \begin{aligned}
M^9=\begin{pmatrix}
        0&0&0&0&0&0&0&0&0&0\\
        0&0&0&0&0&0&0&0&0&0\\
        0&0&0&0&0&0&0&0&0&0\\
        0&0&0&0&0&0&0&0&0&0\\
        0&0&0&0&0&0&0&0&0&0\\
        0&0&0&0&0&0&0&0&0&0\\
        0&0&0&0&0&0&1&1&0&0\\
        0&0&0&0&0&0&1&1&0&0\\
        0&0&0&0&0&0&0&0&1&1\\
        0&0&0&0&0&0&0&0&1&1\\
    
    \end{pmatrix},\;
    M^{10}=\begin{pmatrix}
        0&0&0&0&0&0&0&0&0&0\\
        0&0&0&0&0&0&0&0&0&0\\
        0&0&0&0&0&0&0&0&0&0\\
        0&0&0&0&0&0&0&0&0&0\\
        0&0&0&0&0&0&0&0&0&0\\
        0&0&0&0&0&0&0&0&0&0\\
        0&0&0&0&0&0&0&0&1&1\\
        0&0&0&0&0&0&0&0&1&1\\
        0&0&0&0&0&0&1&1&1&1\\
        0&0&0&0&0&0&1&1&1&1\\

    \end{pmatrix}.
\end{aligned}
\end{equation}
Each of these 10 matrices corresponds to a solution family in \eqref{eq:10solu} via the back modular transformation.
\section{Algebraic aspects of RCFT and TDL}\label{Algebraic aspects of RCFT and TDL}
TDLs represent a natural generalization of boundaries in boundary Conformal Field Theory (CFT) as extended objects within \(2\)D CFTs\cite{Cardy:2004hm}. Consider two CFTs defined in the upper and lower halves of the complex plane, with holomorphic components \(T^1(z)\), \(T^2(z)\), and antiholomorphic components \(\bar{T}^1(\bar{z})\), \(\bar{T}^2(\bar{z})\) of the stress tensor. In this scenario, defect is positioned along the real axis. We say a defect is conformal if 
\begin{equation}
    T^{1}(x)-\bar{T}^1(x) = T^2(x)-\bar{T}^2(x)\quad \forall x\in\R.
\end{equation}
There is an obvious solution that we are interested in 
\begin{equation}
     T^{1}(x)= T^{2}(x);\ \bar{T}^1(x)=\bar{T}^2(x),
\end{equation}
in which case, it is termed as totally transmissive or topological. In this paper, we specifically focus on the set up where the two CFTs are identical. Topological defect lines can be deformed on the complex plane, as long as they do not pass through the field insertion point. This condition is referred to as isotopy invariance\cite{Chang:2018iay}. It is characterized by the following equations:
\begin{align}\label{isotopy invariace}
    [\mathcal{L},T]=0;\ [\mathcal{L},\bar{T}]=0,
\end{align}
where \(\mathcal{L}\) is a topological defect line. Furthermore, correlators within such a theory only depend on the homotopy class of the TDL configuration. 

Before diving into the algebraic structure, we start from emphasising the distinctions between {\it{chiral}} CFTs and {\it{full}} CFTs. Both of them have applications in physics. Chiral CFTs are characterised by the Moore-Seiberg data \(\mathcal{C}\) which is a modular tensor category(MTC) as the representations of the vertex operator algebra \(\mathcal{V}\), i.e. \(\mathcal{C}=\mathcal{R}ep\mathcal{(V)}\). Chiral CFTs can be used to describe phenomena such as fractional quantum Hall effect\cite{Moore:1991ks}.
A Full RCFT on an orientable surface is characterized by a pair \((\mathcal{C}, A)\), where \(A\) is a special symmetric Frobenius algebra in \(\mathcal{C}\). Full CFTs arise as world sheet theories of string theories, in two-dimensional critical phenomena, and in effectively one-dimensional systems\cite{Fuchs:2023ngi}. Mathematically, one of the key distinctions lies in the fact that chiral CFTs are defined on Riemannian surfaces where correlators exhibit multi-valued behavior, whereas full CFTs are defined on conformal real two-manifolds with single-valued correlators. In fact, for full CFTs, what we require are stratified manifolds whose one-dimensional strata are interpreted as defects that separate two (different) phases of CFTs. 

We adopt the notation from the book \cite{Bakalov2000LecturesOT}. Along the reviewing of algebras, modules and bimodules, we will introduce graph calculus for the purpose of intuition and calculation. We work in Abelian category so that we can borrow the manipulations in the category of vector space over \(\CC\) safely. 

A tensor category is a category \(\mathcal{C}\) together with a  bifunctor \(\otimes: \mathcal{C}\times \mathcal{C} \rightarrow C\) with functorial isomorphisms. We require associativity  \(\alpha_{UVW}: (U\otimes V)\otimes W\xrightarrow{\sim}U\otimes (V\otimes W)\) and a tensor identity \(1 \in \Obj(\mathcal{C})\) with natural unit isomorphisms \(\rho_{V}: V\otimes 1\xrightarrow{\sim} V\) and \(\lambda_{V}: 1\otimes V\xrightarrow{\sim} V\) for all \(V\in \Obj(\mathcal{C})\) such that
 \begin{enumerate}
 \item for any \(V_i\in Obj(\mathcal{C})(i=1,2,3,4)\) The diagram
\[\begin{tikzcd}
&((V_{1}\otimes V_{2})\otimes V_{3})\otimes V_{4}\arrow[ld,"\alpha_{1,2,3}\otimes \identity_4",swap]\arrow[rd,"\alpha_{12,2,3}"]&\\
(V_{1}\otimes (V_{2}\otimes V_{3}))\otimes V_{4}\arrow[d,"\alpha_{1,23,4}"]&&(V_{1}\otimes V_{2})\otimes (V_{3}\otimes V_{4})\arrow[d,"\alpha_{1,2,34}"]\\
V_{1}\otimes ((V_{2}\otimes V_{3})\otimes V_{4})\arrow[rr,"\identity_1\otimes\alpha_{2,3,4}"]&&V_{1}\otimes (V_{2}\otimes (V_{3}\otimes V_{4}))
\end{tikzcd}\]
is commutative;
\item for any \(V_1, V_2\in \Obj(\mathcal{C})\) the diagram
\[\begin{tikzcd}
(V_1\otimes 1)\otimes V_2\ar[rr,"\alpha"]\ar[rd,"\identity\otimes\rho",swap]&&V_1\otimes (1\otimes V_2)\ar[ld,"\rho\otimes\identity"]\\
&V_1\otimes V_2&
\end{tikzcd}\]
is commutative.
\end{enumerate}
If the category is braided, there is one more functorial isomorphism,
\begin{align}
c_{VW}:V\otimes W\xrightarrow{\sim} W\otimes V,
\end{align}
and correspondingly the following axiom need to be satisfied:
\begin{enumerate}
\item For any \(V_i\in \Obj(\mathcal{C})(i=1,2,3)\) the diagram
\[
\begin{tikzcd}
&V_{1}\otimes(V_{2}\otimes V_{3})\ar[r,"c_{1,23}"]&(V_{2}\otimes V_{3})\otimes V_{1}\ar[rd,"c_{1,23}"]&\\
(V_{1}\otimes V_{2})\otimes V_{3}\ar[ru,"\alpha_{1,2,3}"]\ar[rd,"c_{1,2}\otimes\identity_{3}",swap]&&&V_{2}\otimes (V_{1}\otimes V_{3})\\
&(V_{2}\otimes V_{1})\otimes V_{3}\ar[r,"\alpha_{2,1,3}"]&V_{2}\otimes (V_{1}\otimes V_{3})\ar[ru,"\identity_{2}\otimes c_{1,3}",swap]&
\end{tikzcd}\]
\item The same diagram but with \(c\) replaced by \(c^{-1}\).
\end{enumerate}
Recall that in the category finite dimensional vector space, there is a dual vector space. A tensor cateogory is rigid if for any \(V\in \Obj(\mathcal{C})\), there exists a right dual with the following two evaluation and co-evaluation morphisms,
\begin{align}
e_{V}&:V^*\otimes V \rightarrow 1\\
i_{V}&:1\rightarrow V\otimes V^*,
\end{align}
such that the following compositions are identity morphisms respectively.
\begin{equation}\label{rigidity1}
V\xrightarrow{i_{V}\otimes \identity_{V}}V\otimes V^*\otimes V \xrightarrow{\identity_{V}\otimes e_{V}} V,
\end{equation}
\begin{equation}\label{rigidity2}
V^*\xrightarrow{\identity_{V}\otimes i_{V}}V^*\otimes V\otimes V^* \xrightarrow{e_V\otimes\identity_{V}} V^*.
\end{equation}
Properties \eqref{rigidity1} and \eqref{rigidity2} are called the rigidity axioms. Similar structures are required in the left dual and we denote the evaluationa and co-evaluation morphism as \(\tilde{e}\) and \(\title{i}\).

If a rigid braided tensor category is a \emph{ribbon category}, there is a functorial isomorphism \(\delta_V: V\xrightarrow{\sim} V^{**}\), such that is satisfy,
\begin{align}
\delta_{V\otimes W} &= \delta_V\otimes\delta_{W}\\
\delta_{1}&=\identity\\
\delta_{V^{*}}&=(\delta_{V}^*)^{-1}
\end{align}
where for \(f^*\in\Hom({V^*,U^*})= \Hom{(U,V)}\).
\begin{definition}
A \emph{modular tensor category} is a ribbon category, which has a simple tensor unit, and a finite number of isomorphism classes of simple objects, namely, \emph{semisimple}. If \(\{U_i| i \in \mathcal{I}\}\) denotes a choice of representatives for these classes, in addition \(s=(s_{ij}\in \End{1})_{i,j\in\mathcal{I}}\) with \(s_{ij} = \Tr{(c_{U_i,U_j}\circ c_{U_j,U_i})}\) is non-degenerate.
\end{definition}

\begin{figure}[htb]
\centering
\begin{tikzpicture}[thick,scale=0.5, every node/.style={scale=0.5}]
\node[scale=2] at (0,0){\(m=\)};
\begin{scope}[shift={(2,0.5)}]
\node at (0,0){\(\bullet\)};
\filldraw[magenta] (0,0) circle (2pt);
\node at (0,1.25){\(A\)};
\node at (1,-1.75){\(A\)};
\node at (-1,-1.75){\(A\)};
\draw[-,thick] (0,1) -- (0,0.1);
\draw[-,thick] (1,-1.5) -- (1,-1) arc (0:86:1);
\draw[-,thick] (-1,-1.5) -- (-1,-1) arc (180:94:1);
\node at (1.5,-0.5){,};
\begin{scope}[shift={(3,0)}];
\node[scale=2] at (0,-0.5){\(\eta=\)};
\begin{scope}[shift={(1,-0.5)}];
\draw[-,thick] (0,0.5) -- (0,-0.46);
\filldraw[cyan] (0,-0.5) circle (2pt);
\node at (0,0.75){\(A\)};
\node at (0,-1){\(1\)};
\node at (0.5,0){,};
\begin{scope}[shift={(1,0.1)}];
\node[scale=2] at (1.25,0){\(\Delta=\)};
\begin{scope}[shift={(1.5,-0.5)}];
\draw[-,thick] (2,-1) -- (2,-0.1);
\filldraw[magenta] (2,0) circle (2pt);
\draw[-,thick] (1,1.5) -- (1,1) arc (180:266:1);
\draw[-,thick] (3,1.5) -- (3,1) arc (0:-86:1);
\node at (2,-1.25){\(A\)};
\node at (1,1.75){\(A\)};
\node at (3,1.75){\(A\)};
\node at (3.5,0.5){,};
\begin{scope}[shift={(5,0.5)}];
\node[scale=2] at (0.5,0){\(\epsilon=\)};
\draw[-,thick] (1.5,-0.5) -- (1.5,0.5);
\filldraw[blue] (1.5,0.5) circle (2pt);
\node at (1.5,1){\(1\)};
\node at (1.5,-0.75){\(A\)};
\end{scope}
\end{scope}
\end{scope}   
\end{scope}
\end{scope}
\end{scope}
\end{tikzpicture}
\caption{Structures of algebra and coalgebra}
\label{algegra and coalgebra}
\end{figure}
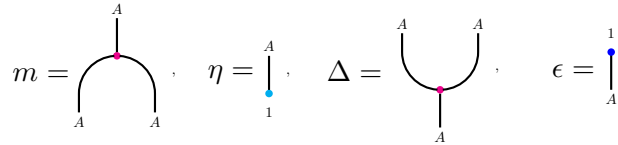

An algebra in \(\mathcal{C}\) is a triple \((A, m, \eta)\), where \(A\) is an object of  \(\mathcal{C}\) equipped with two morphisms \(m:A\otimes A\rightarrow A \) and \(\eta : 1 \rightarrow A\) satisfying the usual associativity and unit properties. Similarly, a coalgebra is an object \(A\), equipped with another two morphisms \(\Delta \in \Hom(A, A\otimes A)\) and \(\epsilon \in \Hom(A, 1)\) that satisfy the usual coassociativity and counit properties. We will use the  graphical representations in figure \ref{algegra and coalgebra} later,

A Frobenius algebra in a tensor category \(\mathcal{C}\) is an object that is both an algebra and a coalgebra. The product \(m\) and coproduct \(\Delta\) satisfy the compatibility requirement that the coproduct is an intertwiner of \(A\)-bimodules, namely
\begin{equation}
(\identity_A\otimes m)\circ (\Delta\otimes \identity_A) = \Delta\otimes m = (m\otimes \identity)\circ (\identity_A\otimes \Delta).
\end{equation}
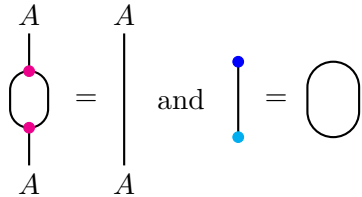
\begin{figure}[ht]
\centering
\begin{tikzpicture}
\draw[-,thick] (0,-0.75) -- (0,-0.25) arc (-90:0:0.25) -- (0.25,0.25) arc (0:180:0.25);
\draw[-,thick] (0,-0.25) arc (-90:-180:0.25) -- (-0.25,0.25);
\draw[-,thick] (0,0.5) -- (0,1);
\node at (0,-1){\(A\)};
\node at (0,1.25){\(A\)};
\filldraw[magenta] (0,-0.25) circle (2pt);
\filldraw[magenta] (0,0.5) circle (2pt);
\node at (0.75,0.125){\(=\)};
\draw[-,thick] (1.25,-0.75) -- (1.25,1);
\node at (1.25,-1){\(A\)};
\node at (1.25,1.25){\(A\)};
\node at (2,0.125){and};
\begin{scope}[shift={(2.75,0.125)}];
\draw[-,thick] (0,-0.5) -- (0,0.5);
\filldraw[cyan] (0,-0.5) circle (2pt);
\filldraw[blue] (0,0.5) circle (2pt);
\node at (0.5,0){\(=\)};
\begin{scope}[shift={(1.25,-0.16)}];

\draw[-,thick] (-0.34,0) -- (-0.34,0.32) arc (-180:-360:0.34) -- (0.34,0) arc (0:-180:0.34);
\end{scope}
\end{scope}
\end{tikzpicture}
\caption{Special Frobenius algebra.}
\label{special}
\end{figure}

\begin{figure}[ht]
\centering
\begin{tikzpicture}
\draw[thick] (-0.75,-0.25) -- (-0.75,-0.75);
\draw[->, thick] (-1.5,0) -- (-1.5,-1) arc (180:270:0.25);
\draw[-,thick]  (-1.25,-1.25) arc (270:360:0.25);
\draw[-,thick] (-1,-1) arc (180:0:0.25) -- (-0.5, -1.75);
\node at (-1.25,-1.25){\(>\)};
\node at (-0.5, -2){$A$};
\node at (-1.5, 0.25){$A^*$};
\filldraw[magenta] (-0.75,-0.75) circle (2pt);
\filldraw[blue] (-0.75,-0.25) circle (2pt);
\node at (0,-0.75){\(=\)};
\draw[->, thick] (1.5,0) -- (1.5,-1) arc (0:-90:0.25);
\node at (1.25,-1.25){\(<\)};
\draw[-,thick]  (1.25,-1.25) arc (-90:-180:0.25);
\draw[-,thick] (1,-1) arc (0:180:0.25) -- (0.5, -1.75);
\draw[thick] (0.75,-0.25) -- (0.75,-0.75);
\filldraw[magenta] (0.75,-0.75) circle (2pt);
\filldraw[blue] (0.75,-0.25) circle (2pt);
\node at (0.5,-2){$A$};
\node at (1.5,0.25){$A^*$};
\end{tikzpicture}
\caption{Symmetric Frobenius algebra}
\label{symmetric}
\end{figure}
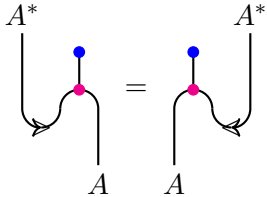

A Frobenius algebra is called 
\begin{itemize}
\item special (figure \ref{special}) if \(m\circ \Delta = \beta_1 \identity_A\) and \(\epsilon\circ \eta = \beta_2\identity_{1}\) for non zero constants \(\beta_1,\beta_2\in\CC^{*}\).

\item symmetric (figure \ref{symmetric}) if \( [\identity_{A^*} \otimes (\epsilon\circ m)]\circ(\tilde{i}_A\otimes\identity_A)=[(\epsilon\circ m)\otimes \identity_{A^*}]\circ (\identity_A \otimes i_A) \)
\end{itemize}

The reason why we define such a Frobenius algebra structure is not {\it a priori} obvious. As we will see later, it turns out this kind of structure serves as an crucial ingredient in our construction. The original idea of formulating Frobenius Algebra is borrowed from lattice topological field theory in two dimensions, and just like lattice TFT, Morita equivalent algebras give rise to equivalent theories\cite{Fukuma:1993hy}. In the vertex operator algebra community, \(A\) is known as an open-string vertex algebra with a non-degenerate invariant bilinear form\cite{Huang:2003ju,Kong:2008urc,kong2011conformal}. We next sketch a more physical point of view.

Consider a boundary CFT, one can regard the OPE between bulk fields and  boundary fields as the action of \(\mathcal{C}\) on the boundary fields, 
\begin{equation}
\phi_{i}(z)\phi_{j}(x) = \sum_{k}c_{ij}^{k}(x,z) \phi_{k}(x).
\end{equation}
We require the action to be associative,
\begin{equation}
\left(\phi_{i_1}(z_1)\phi_{i_2}(z_2)\right)\phi_{j}(x)=\phi_{i_1}(z_1)\left(\phi_{i_2}(z_2)\phi_{j}(x)\right),
\end{equation}
where \(x\) and \(z\) are coordinates on the boundary and bulk respectively. We thus find that this is exactly the structure needed to define a module category for a MTC \(\mathcal{C}\). If the boundary data is encoded in a module category \(\mathcal{M}\), then it's proved in \cite{ostrik2001module} that \(\mathcal{M}\) is equivalent to the category \(\mathcal{C}_A\) of (left) \(A\)-modules in \(\mathcal{C}\) for some algebra \(A\) in \(\mathcal{C}\). Similarly  \(\mathcal{C_M^*}\) is  equivalent to the category \(\mathcal{C_{A|A}}\) of \(A\)-bimodules in \(\mathcal{C}\), where \(\mathcal{C_M^*} = \mathcal{F}un_\mathcal{C}\mathcal{(M,M)}\) are module endofunctors. We will introduce the concepts of \(A\)-(bi)module momentarily. A thoroughly computations on three-state Potts model that taking the module category point of view (instead of employing Frobenius algebra) has been done in \cite{Vanhove:2021nav}, and the twisted partition functions are the same with \cite{Frohlich:2006ch}. This agreement provides substantial support for the idea that Frobenius algebras can indeed assist in the construction of RCFT.   

It's important to note that there can be more than one Frobenius algebras in a given MTC. For example, consider the tetra-critical Ising model and the three-state Potts model. Despite sharing the same MTC, denoted as \(\mathcal{C}_{5,6} = \mathfrak{su}(2)_4 \times \text{Lee-Yang}\), they have distinct Frobenius algebras \(1\) and \(1\oplus w\). As we mentioned before, theories are defined on world sheets and they can be separated by defects. We thus say theories separated by a defect are in different phases if their Frobenius algebras are different. A special type of such defect is that it has the same phase \(A\) on both sides. The aforementioned topological defect line is in such case and they commute with the energy momentum tensor in addition.

\begin{figure}[htb]
\centering
\begin{tikzpicture}
\draw[-,thick] (0,-1)--(0,1);
\draw[-,thick] (-0.5,-1) -- (-0.5,-0.5) arc (-180:-270:0.5);
\draw[-,thick] (0.5,-1) -- (0.5,0) arc (0:90:0.5);
\node at (-0.5,-1.25){\(A\)};
\node at (0.5,-1.25){\(A\)};
\node at (0,-1.25){\(\dot{X}\)};
\node at (0,1.25){\(\dot{X}\)};
\node at (-0.75,-0.5){\(\rho_{\ell}\)};
\node at (0.75,0){\(\rho_r\)};
\end{tikzpicture}
\caption{Left and right module structures.}
\label{left and right module}
\end{figure}
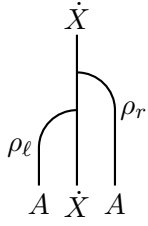

The mathematical structure for the defect is an \(A\)-bimodule, and it consists of a triple \(X = (\dot{X}, \rho_{\ell}, \rho_r)\), where \(\dot{X}\) is an object in the category \(\mathcal{C}\), \(\rho_{\ell}\) is an element of \(\Hom(A \otimes \dot{X},\dot{X})\) and \(\rho_r\) is an element of \(\Hom(\dot{X}\otimes A,\dot{X}\)). The graphical representation is figure \ref{left and right module}.
These elements are subject to the following equalities:
    \begin{itemize}
    \item Unit property: \(\rho_{\ell} \circ\left(\eta_A \otimes \identity_{\dot{X}}\right)=\identity_{\dot{X}}\) and \(\rho_r \circ\left(\identity_{\dot{X}} \otimes \eta_A\right)=\identity_{\dot{X}}\)
    \item Representation property: \(\rho_{\ell} \circ\left(m_A \otimes \identity_{\dot{X}}\right)=\rho_{\ell} \circ\left(\identity_A \otimes \rho_{\ell}\right)\) and \(\rho_r \circ\left(\identity_{\dot{X}} \otimes m_A\right)=\rho_r \circ\left(\rho_r \otimes \identity_A\right)\).
    for example in the left module case, pictorially,
    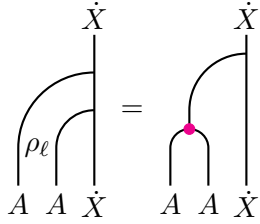
\begin{figure}[htb]
\centering
\begin{tikzpicture}
\draw[-,thick] (0,-1)--(0,1);
\draw[-,thick] (-0.5,-1) -- (-0.5,-0.5) arc (-180:-270:0.5);
\draw[-,thick] (-1,-1) -- (-1,-0.5) arc (-180:-270:1);
\node at (-0.5,-1.25){\(A\)};
\node at (-1,-1.25){\(A\)};
\node at (0,-1.25){\(\dot{X}\)};
\node at (0,1.25){\(\dot{X}\)};
\node at (-0.75,-0.5){\(\rho_{\ell}\)};
\node at (0.5,0){\(=\)};
\draw[-,thick] (1,-1) -- (1,-0.5) arc (-180:-360:0.25) -- (1.5,-1);
\draw[-,thick] (1.25,-0.25) -- (1.25,0) arc (-180:-270:0.75);
\draw[-,thick] (2,-1) -- (2,1);
\node at (1,-1.25){\(A\)};
\node at (1.5,-1.25){\(A\)};
\node at (2,-1.25){\(\dot{X}\)};
\node at (2,1.25){\(\dot{X}\)};
\filldraw[magenta] (1.25,-0.25) circle (2pt);

\end{tikzpicture}
\caption{Representation property of Module}
\label{representation property(left)}
\end{figure}

    \item Left and right action commute: \(\rho_{\ell} \circ\left(\identity_A \otimes \rho_r\right)=\rho_r \circ\left(\rho_{\ell} \otimes \identity_A\right)\)
    \end{itemize}
The morphism between two \(A\)-bimodules i.e. the intertwiners, are given by the a subset of  morphism in \(\mathcal{C}\) that compatible with \(\rho_{l,r}\)
\begin{equation}
\Hom_{\mathcal{C}_{A|A}}(\dot{X},\dot{Y}) = \{f\in \Hom(\dot{X},\dot{Y})| f\circ \rho_{\ell}=\rho_{\ell}\circ(\identity_A\otimes f), f\circ \rho_r=\rho_r\circ (f\otimes \identity_A)\}.
\end{equation}
The tensor/fusion product of two defects are then defined by the image of the morphism \(P_{X,Y}\in\Hom{(X\otimes Y,X\otimes Y)}\).
\begin{equation}
P_{X,Y}\coloneqq (\rho_X\otimes \rho_Y)\circ (\identity\otimes(\Delta\circ\eta)\otimes\identity)
\end{equation}
\begin{equation*}
\begin{tikzpicture}
\node at (0,0){\(P_{X,Y}\, = \)};
\draw[-,thick] (1,-1) -- (1,1);
\draw[-,thick] (2,-1) -- (2,1);
\draw[-,thick] (1.5,-0.5) -- (1.5,0) arc (-180:-270:0.5);
\draw[-,thick] (1.5,0) arc (0:90:0.5);
\filldraw[magenta] (1.5,0) circle (2pt);
\filldraw[blue] (1.5,-0.5) circle (2pt);
\end{tikzpicture}  
\end{equation*}
that restrict \(X\otimes Y\in \mathcal{C}\) to \(X\otimes_{A}Y\in \mathcal{C}_{A|A}\) We would also like to highlight that there is a braiding structure for defects inherited from \(\mathcal{C}\) dressed by some module structure. Although braiding between defects is impossible since they are restricted to the \(2\)D manifold, the braiding between defects and objects in \(\mathcal{C}\) is possible because the latter represents the ribbon in the \(3\)D manifold within the TFT construction. Additionally, some Verlinde-like formulas have been considered in \cite{deshpande2018twisted,Shen:2019wop,deshpande2022crossed}. We aim to determine the non-negative integer matrix in Section \ref{defct Cardy condition} with the help of a certain Verlinde-like formula in the future. 

Given an \(A\)-bimodule \(X = (\dot{X}, \rho_{\ell}, \rho_r)\) in a ribbon category \(\mathcal{C}\), and \(U,V\in \Obj\mathcal{C}\), we can use the braiding to define several bimodule structures on \(U\otimes\dot{X}\otimes V\in \Obj\mathcal{C}\). A typical example we will use is
\begin{align}
&U\otimes^+X\otimes^-V\coloneqq\nonumber\\
&(U\otimes\dot{X}\otimes V, (\identity_U\otimes\rho_{\ell}\otimes\identity_V)\circ (c^{-1}_{U,A}\otimes \identity_{\dot{X}}\otimes  \identity_ \identity),( \identity_U\otimes\rho_r \identity_ V)\circ(\identity_U\otimes\ \identity_{\dot{X}}\otimes c^{-1}_{V,A})).
\end{align}

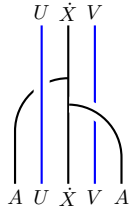
\begin{figure}[htb]
\centering
\begin{tikzpicture}[scale=0.7, every node/.style={scale=0.7}]
\draw[-,thick] (0,0)--(0,3);
\draw[-,thick,blue] (0.5,0) -- (0.5,3);
\draw[-,thick] (-1,0) -- ((-1,1) arc (-180:-270:1);
\draw[-,thick,cross,blue] (-0.5,0) -- (-0.5,3);
\draw[-,thick,cross] (1,0) -- (1,0.5) arc (0:89:1);
\node at (-1,-0.25){\(A\)};
\node at (-0.5,-0.25){\(U\)};
\node at (0,-0.25){\(\dot{X}\)};
\node at (0.5,-0.25){\(V\)};
\node at (1,-0.25){\(A\)};
\node at (-0.5,3.25){\(U\)};
\node at (0,3.25){\(\dot{X}\)};
\node at (0.5,3.25){\(V\)};
\end{tikzpicture}
\caption{Define bimodule structures using braiding.}
\label{induced bimodule by braiding}
\end{figure}
\noindent This kind of bimodule and its variations are related to the CFT quantities. A summary is in table \ref{dictionary between CFT quantity and algebraic notion}.
The connection between TFTs and RCFTs originates from the seminal work of \cite{Witten:1988hf}, which established the equivalence between the wave functional of Chern-Simons theory and the conformal block of the WZW model. Over the past three decades, numerous generalizations have been done in condensed matter physics, high-energy physics, and mathematical physics. One of the most exciting conclusions is that the \((2+1)\)D bulk phase \(\mathcal{D}\) for a given \((1+1)\)D boundary \(\mathcal{C}\) is uniquely determined by the Drinfeld center of \(\mathcal{C}\) \cite{Vanhove:2018wlb, Kong_2018, Kong_20182, Kong:2019cuu, Lootens:2019ghu, Ji:2019jhk, Kong_2020, Gaiotto:2020iye, Bhardwaj:2020ymp, Apruzzi:2021nmk, Kong:2021equ, Kong_20202, Kong:2020iek, Kong:2020jne, albert2021spin, liu2022noninvertible, Freed_2022, Xu:2022rtj, Chen:2022wvy, Moradi:2022lqp, Chatterjee_2023, Lootens_2023, Chatterjee_2023B}.

\bibliographystyle{unsrt}
\bibliography{ref}


\end{document}